\documentclass{aa}
\usepackage[varg]{txfonts}

\usepackage{placeins}
\usepackage[colorlinks=true, linkcolor=blue, urlcolor=blue, citecolor=blue, anchorcolor=blue]{hyperref}
            
\usepackage{natbib}
\bibpunct{(}{)}{;}{a}{}{,}

\usepackage{orcidlink}

\title{The ESO SupJup Survey} \subtitle{IX. Isotopic evidence of a recent formation for Luhman 16AB}
\titlerunning{Isotopic evidence of a recent formation for Luhman 16AB}

\author{
S. de Regt\inst{\ref{instLeiden}}\orcidlink{0000-0003-4760-6168} \and
I. A. G. Snellen\inst{\ref{instLeiden}}\orcidlink{0000-0003-1624-3667} \and
D. Gonz\'alez Picos\inst{\ref{instLeiden}}\orcidlink{0000-0001-9282-9462} \and
S. Gandhi\inst{\ref{instWarwick},\ref{instCEH}}\orcidlink{0000-0001-9552-3709} \and
N. Grasser\inst{\ref{instLeiden}}\orcidlink{0009-0009-6634-1741} \and
A. Y. Kesseli\inst{\ref{instIPAC}}\orcidlink{0000-0002-3239-5989} \and
R. Landman\inst{\ref{instLeiden}}\orcidlink{0000-0002-7261-8083} \and
P. Molli\`ere\inst{\ref{instMPIA}}\orcidlink{0000-0003-4096-7067} \and
E. Nasedkin\inst{\ref{instTCD}}\orcidlink{0000-0002-9792-3121} \and
T. Stolker\inst{\ref{instLeiden}}\orcidlink{0000-0002-5823-3072} \and
Y. Zhang\inst{\ref{instCalTech}}\orcidlink{0000-0003-0097-4414}
}

\institute{
Leiden Observatory, Leiden University, P.O. Box 9513, 2300 RA, Leiden, The Netherlands\\\email{regt@strw.leidenuniv.nl} \label{instLeiden} \and
Department of Physics, University of Warwick, Coventry CV4 7AL, UK \label{instWarwick} \and
Centre for Exoplanets and Habitability, University of Warwick, Gibbet Hill Road, Coventry CV4 7AL, UK \label{instCEH} \and
IPAC, Mail Code 100-22, Caltech, 1200 E. California Boulevard, Pasadena, CA 91125, USA \label{instIPAC} \and
Max-Planck-Institut für Astronomie, Königstuhl 17, 69117 Heidelberg, Germany \label{instMPIA} \and
School of Physics, Trinity College Dublin, University of Dublin, Dublin 2, Ireland \label{instTCD} \and
Department of Astronomy, California Institute of Technology, Pasadena, CA 91125, USA \label{instCalTech}
}

\date{Received date / Accepted date}

\abstract
{The distinct formation pathways proposed for directly imaged exoplanets and isolated brown dwarfs might leave imprints in the inherited chemical composition. Elemental and isotopic tracers could help inform the suspected histories, but this requires a careful characterisation of the sub-stellar atmospheres. In particular, objects at the L-T transition exhibit signs of dynamics that can drive their atmospheres out of chemical equilibrium.}
{In this work, we studied the nearest L-T brown dwarfs, Luhman 16A and B, to assess the chemical disequilibrium in their atmospheres. We also investigated the elemental and isotopic compositions in the context of their probable formation history within the Oceanus moving group.} 
{As part of the ESO SupJup Survey, we obtained spatially resolved CRIRES$^+$ K-band spectra of the binary. These high-resolution observations were analysed using an atmospheric retrieval framework that couples the radiative transfer code \texttt{petitRADTRANS} to the \texttt{MultiNest} sampling algorithm.}
{We detect and retrieve the abundances of $^{12}$CO, H$_2$O, CH$_4$, NH$_3$, H$_2$S, HF, and the $^{13}$CO isotopologue. We find that both atmospheres are in chemical disequilibrium with somewhat stronger vertical mixing in Luhman 16A compared to B ($K_\mathrm{zz,A}\sim$\,$10^{8.7}$ vs $K_\mathrm{zz,B}\sim$\,$10^{8.2}\ \mathrm{cm^2\ s^{-1}}$). The tested chemical models, free-equilibrium and disequilibrium chemistry, yield consistent mixing ratios and agree with earlier work at shorter wavelengths. The free-chemistry gaseous $\mathrm{C/O}$ ratios show evidence of oxygen trapping in silicate-oxide clouds. While the $\mathrm{C/O}$ ratios are consistent with the solar composition, the metallicities are modestly enhanced with $\mathrm{[C/H]}\sim$\,$0.15$. The carbon isotope ratios are measured at $\mathrm{^{12}C/^{13}C_A}=74^{+2}_{-2}$ and $\mathrm{^{12}C/^{13}C_B}=74^{+3}_{-3}$.}
{The coincident constraints of metallicities and isotopes across the binary provide further evidence in favour of a common formation. The $\mathrm{^{12}C/^{13}C}$ ratios are aligned with the present-day interstellar medium but lower than the Solar System value. This suggests a recent inheritance and corroborates the relatively young age ($\sim$\,$500\ \mathrm{Myr}$) of Luhman 16A and B as members of the Oceanus moving group.} 


\usepackage{subcaption}
\usepackage{float}
\usepackage{subfiles}
\usepackage{pdflscape}
\usepackage{xcolor}

\begin{document}

\maketitle

\section{Introduction}
Brown dwarfs at the transition between spectral types L and T go through a remarkable change in their spectro-photometric appearance \citep{Kirkpatrick_2005}. The cooler T dwarfs exhibit a shift towards bluer colours, which has often been associated with the dissipation of photospheric clouds (e.g. \citealt{Burrows_ea_2006,Saumon_ea_2008}) as well as thermo-chemical instabilities \citep{Tremblin_ea_2016}. In this temperature regime ($1200$--$1400\ \mathrm{K}$; \citealt{Faherty_ea_2014}), the sub-stellar objects also experience a transition in their atmospheric chemistry. The higher L-dwarf temperatures favour CO as the main carbon-bearing species, while the T-type dwarfs are defined by the strengthening CH$_4$ absorption in their spectra \citep{Cushing_ea_2005,Kirkpatrick_2005}. The observables are further impacted by atmospheric dynamics that likely drive the photometric variability that is commonly seen for L-T objects \citep{Radigan_ea_2014}. In addition, strong vertical mixing within the atmosphere inhibits the conversion of CO into CH$_4$ at low pressures \citep{Hubeny_ea_2007,Zahnle_ea_2014,Mukherjee_ea_2022}. The resulting chemical disequilibrium can be assessed through measurements of under-abundant CH$_4$ or enhanced CO absorption (e.g. \citealt{Noll_ea_1997,Geballe_ea_2009,Miles_ea_2020}).

Directly imaged exoplanets and isolated brown dwarfs share similar atmospheric temperatures and, by extension, are in comparable chemical states \citep{Faherty_ea_2016}. The two populations likely have distinct formation histories, resulting in differences in the accreted elemental abundances. The composition of giant exoplanets is thought to be imprinted with the chemical and isotopic fractionation that occurs at various radii throughout the circumstellar disc \citep{Oberg_ea_2021,Bergin_ea_2024}. Isolated brown dwarfs, on the other hand, form via gravitational collapse, and their compositions are expected to reflect the chemical content of the parent molecular cloud \citep{Bate_ea_2002,Chabrier_ea_2014}. Elemental abundance ratios such as the $\mathrm{C/O}$, $\mathrm{N/O}$, and refractory-to-volatile ratios have been suggested as valuable planet-formation tracers \citep{Oberg_ea_2011,Madhusudhan_ea_2012,Turrini_ea_2021,Lothringer_ea_2021}, but isotope ratios have been proposed as complementary tracers as well \citep{Molliere_ea_2019a,Morley_ea_2019}. The carbon isotope ratio, $\mathrm{^{12}C/^{13}C}$, may be particularly useful for assessing the formation epochs of brown dwarfs and low-mass stars. Owing to their fully convective interiors and lack of carbon production, their atmospheres preserve the $\mathrm{^{12}C/^{13}C}$ ratio of their natal environments. While stars ultimately expel both isotopes back into the interstellar medium (ISM), they have separate nucleosynthetic origins, with $^{12}$C synthesised in the triple-$\alpha$ process and $^{13}$C created as a secondary element through the CNO cycle \citep{Wiescher_ea_2010}. As such, the $\mathrm{^{12}C/^{13}C}$ ratio is expected to decrease over time as the necessary metals for the CNO cycle become accessible \citep{Romano_ea_2022}. In agreement with predictions of galactic chemical evolution, \citet{Gonzalez_Picos_ea_2025b} find that nearby M-dwarf stars with increased metallicities are also progressively enriched in $^{13}$C. 

The WISE J104915.57-531906.1 system, or Luhman 16, is the nearest brown dwarf binary, at a distance of $\sim$\,$2\ \mathrm{pc}$ \citep{Luhman_2013,Bedin_ea_2024}. The primary and secondary components, A and B, are classified, respectively, as L7.5 and T0.5$\pm$1.0 \citep{Burgasser_ea_2013}. Coincidentally, the kinematic motion of Luhman 16AB yields an association with Oceanus, a moving group with an estimated age of $\sim$\,$500\ \mathrm{Myr}$ \citep{Gagne_ea_2023}. The high apparent brightnesses of the two objects have led to numerous observations, which in turn have been used to characterise their orbits (e.g. \citealt{Lazorenko_ea_2018,Bedin_ea_2024}), polarisation \citep{Millar-Blanchaer_ea_2020}, variabilities (e.g. \citealt{Biller_ea_2013,Biller_ea_2024,Buenzli_ea_2015a,Fuda_ea_2024,Chen_ea_2025}), and spectral energy distributions (e.g. \citealt{Faherty_ea_2014,Crossfield_ea_2014,Lodieu_ea_2015,Kellogg_ea_2017,Chen_ea_2025,Ishikawa_ea_2025}). Given their status as benchmark L-T objects, Luhman 16A and B are well-suited targets for studying the presence of chemical disequilibria. Additionally, measurements of elemental and isotopic compositions can provide further constraints on the suspected formation of the binary. 

In this paper we present an analysis of the atmospheric chemistry of Luhman 16AB using high-resolution K-band spectra. Section~\ref{sect:methods} describes the reduction of our observations and the modelling framework that was used to infer the atmospheric properties. Section~\ref{sect:results} outlines the results, and Sect.~\ref{sect:discussion} discusses them in the context of previous studies as well as the implied formation history of Luhman 16AB. In Sect.~\ref{sect:conclusions} we summarise the conclusions drawn from this atmospheric retrieval analysis. 

\section{Methods} \label{sect:methods}
\subsection{Observations and reduction}
We observed the Luhman 16 binary as part of the ESO SupJup Survey (programme ID: 110.23RW; see \citealt{de_Regt_ea_2024}) on January 1, 2023, using the CRyogenic high-resolution InfraRed Echelle Spectrograph (CRIRES$^+$; \citealt{Dorn_ea_2023}). The K2166 wavelength setting was employed to ensure coverage of the $^{13}$CO ($\nu=2-0$) band head. A single nodding cycle ($4\times 300\ \mathrm{s}$) with the $0.4"$ slit resulted in a signal-to-noise of $400$ and $300$ (at $\sim$\,$2345\ \mathrm{nm}$), respectively for Luhman 16A and B. These K-band observations were taken prior to the J-band spectra studied in \citet{de_Regt_ea_2025}. Following a similar reduction procedure using \texttt{excalibuhr}\footnote{\url{https://github.com/yapenzhang/excalibuhr}} \citep{Zhang_ea_2024}, we extracted the spectra of each component separately. The optical seeing of $\sim$\,$0.8"$ and a projected separation of $\sim$\,$0.81"$ \citep{Bedin_ea_2024} resulted in a degree of blending of the two spectral traces. We applied a correction for the contamination in the 12-pixel wide extraction apertures, following Appendix A of \citet{de_Regt_ea_2025}. Our K-band observations from the preceding night, December 31, 2022, suffered from worse seeing ($\sim$\,$1.5"$) where the binary components could not be spatially resolved. While it is possible to separate them via their radial and rotational velocities, inferring distinct atmospheric properties proved challenging, and we therefore only studied the observations from the second night.

The standard star HD 93563 was observed with the same settings and reduced in a similar manner. We applied \texttt{molecfit} \citep{Smette_ea_2015} to the standard star spectrum to model the telluric transmission. The obtained model is subsequently used to correct the telluric absorption lines in the Luhman 16AB spectra. We masked out pixels with a transmissivity below $\mathcal{T}<80\%$, as the deepest tellurics are generally poorly fit. \texttt{Molecfit} used first-degree polynomials to fit the continuum, which we divided by the Planck curve corresponding to the standard star ($T_\mathrm{eff}=15\,000\ \mathrm{K}$; \citealt{Arcos_ea_2018}). The Luhman 16 spectra were subsequently corrected for this wavelength-dependent throughput. Since \texttt{molecfit} also fits the telluric line-widths, we obtained an estimated resolution of $R\sim60\,000$ ($\sim$\,$5\ \mathrm{km\ s^{-1}}$), which we used to apply instrumental broadening to the model spectra. 

\begin{figure*}[h!]
    \centering
    \includegraphics[width=17cm]{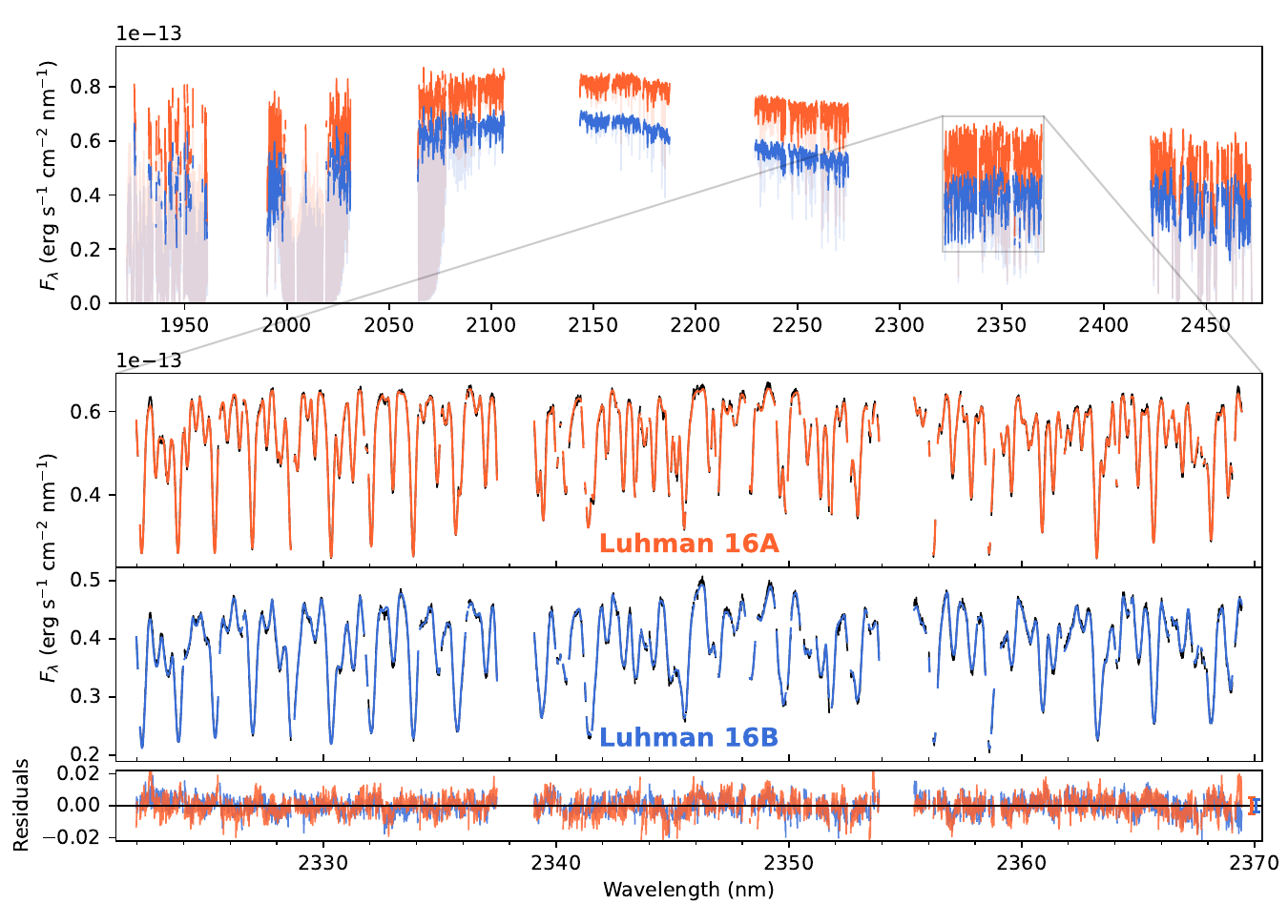}
    \caption{CRIRES$^+$ K-band spectra of Luhman 16A and B in orange and blue, respectively. \textit{Top panel}: Seven spectral orders covered in the K2166 wavelength setting. The telluric absorption is shown as transparent lines. \textit{Lower panels}: Zoom-in of the sixth order. The black observed spectra are overlaid with the best-fitting free-chemistry models in orange and blue. The mean scaled uncertainties are displayed to the right of the residuals in the bottom panel. The fits to the other spectral orders can be found in Appendix~\ref{app:best_fitting_spectra}.}
    \label{fig:spectrum}
\end{figure*}

For plotting purposes, we flux-calibrated the Luhman 16 spectra to match the broadband photometry ($\mathrm{K_A}=9.44\pm0.07$ and $\mathrm{K_B}=9.73\pm0.09\ \mathrm{mag}$; \citealt{Burgasser_ea_2013}). Figure~\ref{fig:spectrum} shows the calibrated spectra of Luhman 16A and B. The high spectral resolution gives a somewhat noisy appearance in the upper panel, but the zoomed-in panels highlight the superb quality of these observations. The K2166 wavelength setting covers seven spectral orders, measured on three detectors, resulting in $7\cdot3=21$ order-detector pairs, or chips. The upper panel of Fig.~\ref{fig:spectrum} shows that the bluest chips have few usable pixels remaining after the telluric masking. 

\subsection{Retrieval framework}
We employed a retrieval framework to constrain the atmospheric properties of the Luhman 16 binary. The radiative transfer code \texttt{petitRADTRANS} (\texttt{pRT}; version 3.1; \citealt{Molliere_ea_2019,Molliere_ea_2020,Blain_ea_2024}) is used to generate emission spectra from a set of variables and parameterisations, which are sampled with the \texttt{PyMultiNest} nested sampling algorithm \citep{Feroz_ea_2009,Buchner_ea_2014}. We used 1000 live points at a constant sampling efficiency of $5\%$ to run the retrievals. The retrieved parameters and their priors are summarised in Table~\ref{tab:params}.

\subsubsection{Likelihood and covariance}
Following \citet{de_Regt_ea_2025}, we defined the likelihood as
\begin{align}
    \ln{\mathcal{L}} &= -\frac{1}{2} \sum_i \left(N_i\ln\left(2\pi s_i^2\right)+\ln\left(|\vec{\Sigma}_{0,i}|\right)+\frac{1}{s_i^2}\vec{r}_i^T\vec{\Sigma}_{0,i}^{-1}\vec{r}_i\right), 
\end{align}
where the sum is performed over 21 order-detector pairs $i$ that have $N_i$ valid pixels ($2048$ at most). For each chip, $\vec{\Sigma}_{0,i}$ is the un-scaled covariance matrix and $\vec{r}_i$ is the residual between the data $\vec{d}_i$ and model $\vec{m}_i$, calculated as $\vec{r}_i = \vec{d}_i - \phi_i \vec{m}_i$. The flux- and covariance-scaling parameters, $\phi_i$ and $s_i^2$, are optimised at each model evaluation using Eqs. 4 and 5 of \citet{de_Regt_ea_2025}. Similarly, we modelled correlated noise using a Gaussian process amplitude, $a$, and length-scale, $\ell$ (Eq. 6 of \citealt{de_Regt_ea_2025}).

\subsubsection{Surface gravity, temperature profile, and clouds}
The well-constrained dynamical masses of $M_\mathrm{A}=35.4\pm0.2$ and $M_\mathrm{B}=29.4\pm0.2\ M_\mathrm{Jup}$ \citep{Bedin_ea_2024} were used as prior information. Combined with radii estimates of $R=1.0\pm0.1\ R_\mathrm{Jup}$ \citep{Biller_ea_2024}, this results in Gaussian priors of $\log\textit{g}_\mathrm{A}=4.96\pm0.09$ and $\log\textit{g}_\mathrm{B}=4.88\pm0.09$ on the surface gravities.

We adopted the gradient-based parameterisation of the temperature profile described in \citet{de_Regt_ea_2025} and introduced by \citet{Zhang_ea_2023}. This parameterisation is flexible but also produces physically reasonable profiles by way of the chosen priors for the temperature gradients, $\nabla_i$ (see Table~\ref{tab:params}). 

As in \citet{de_Regt_ea_2025}, we used a grey-cloud parameterisation with an opacity $\kappa_\mathrm{cl,0}$ at a pressure $P_\mathrm{cl,0}$. Above this cloud base, the opacity decays with the power $f_\mathrm{sed}$. 

\subsubsection{Chemistry} \label{sect:meth_chem}
We compared two approaches to modelling the atmospheric chemistries. Firstly, in the free-chemistry approach, we fitted for the abundance of each included species separately. These volume-mixing ratios (VMRs) were assumed to be vertically constant to minimise the number of free parameters. This is a reasonable assumption for this work, as the species encountered in the K band generally do not show a strong depletion through rainout and vertical mixing also helps homogenise their abundances. We included molecular opacity from $^{12}$CO, $^{13}$CO, C$^{18}$O, C$^{17}$O \citep{Li_ea_2015}, H$_2$O \citep{Polyansky_ea_2018}, H$_2^{18}$O, H$_2^{17}$O \citep{Polyansky_ea_2017}, CH$_4$ \citep{Yurchenko_ea_2024}, CO$_2$ \citep{Hargreaves_ea_2025}, NH$_3$ \citep{Al_Derzi_ea_2015, Coles_ea_2019}, HCN \citep{Harris_ea_2006,Barber_ea_2014}, H$_2$S \citep{Azzam_ea_2016, Chubb_ea_2018}, and HF \citep{Li_ea_2013, Coxon_ea_2015, Somogyi_ea_2021}, calculated with \texttt{pyROX}\footnote{\url{https://py-rox.readthedocs.io}} \citep{de_Regt_ea_2025b}. To clarify, we refer to the main isotope when the mass number superscript is omitted. 

In the second model, we simulated atmospheres in chemical disequilibrium using the \texttt{FastChem Cond} code \citep{Kitzmann_ea_2024}. Equilibrium-chemistry calculations are generally not fast enough to be performed for each sample during a retrieval, especially when accounting for condensation. However, reducing the considered gases and condensates to only those with appreciable abundances (Table~\ref{tab:fastchem_reduced}) can bring computation times down to a few tenths of a second with a negligible loss of accuracy, as demonstrated in Appendix~\ref{app:fastchem_validation}. This model allowed us to compute gaseous VMRs on the fly and include rainout condensation, and we could directly fit for the elemental abundances via $\mathrm{[C/H]}$, $\mathrm{[O/H]}$, $\mathrm{[N/H]}$, $\mathrm{[S/H]}$, and $\mathrm{[F/H]}$ (relative to the solar composition of \citealt{Asplund_ea_2021}). Five isotopologue ratios are retrieved to derive the abundances of $^{13}$CO, C$^{18}$O, C$^{17}$O, H$_2^{18}$O, and H$_2^{17}$O with respect to the main isotopologue. Finally, the abundances of H$_2$O, CH$_4$, CO, NH$_3$, HCN, CO$_2$, and their isotopologues are quenched, or held constant, at $P<P_\mathrm{quench}$. These quench points are located where the vertical-mixing timescale, $t_\mathrm{mix}$, is equal to the chemical timescale of the relevant reaction network (e.g. $t_\mathrm{CO-CH_4-H_2O}$), which we adopted from \citet{Zahnle_ea_2014}. The mixing timescale is calculated as
\begin{align}
    t_\mathrm{mix} &= \frac{L^2}{K_\mathrm{zz}} = \frac{(\alpha H)^2}{K_\mathrm{zz}} = \frac{\alpha^2}{K_\mathrm{zz}}\left(\frac{k_B T}{\mu m_\mathrm{p} \textit{g}}\right)^2, 
\end{align}
where the length scale, $L$, is defined as the product of the scale height, $H$, and a factor, $\alpha$, allowing mixing lengths to be shorter than the scale height \citep{Smith_1998, Ackerman_ea_2001}. In this work, we assumed $\alpha=1$ so that any overestimation of the length scale translates into the vertical eddy diffusion coefficient, $K_\mathrm{zz}$, which is retrieved as a free parameter. The mixing timescale further depends on the temperature profile, mean molecular weight, and surface gravity, which also vary with each evaluated model. 

\section{Results} \label{sect:results}
The retrieval model provides a good fit to the spectra of Luhman 16AB, as is demonstrated in the lower panels of Fig.~\ref{fig:spectrum}. The spectral fits to the remaining six orders can be found in Appendix~\ref{app:best_fitting_spectra}. In general, the residuals are contained within the expected uncertainties (indicated as error bars), with the exception of telluric-dominated wavelengths at the longest K-band wavelengths. 

\subsection{Detection of species} \label{sect:detections}
We confirmed the presence of H$_2$O and $^{12}$CO from a visual inspection of the absorption features (i.e. Fig.~\ref{fig:spectrum}). A more detailed investigation was required to report detections of the less abundant species. Hence, we carried out a cross-correlation analysis between the residuals of a model without species $X$ ($\vec{m}_{i,\mathrm{w/o}\ X}$) and a template of the contribution from $X$ ($\vec{m}_{i}-\vec{m}_{i,\mathrm{w/o}\ X}$). The cross-correlation coefficient at a velocity $\textit{v}$ was then summed over the chips via
\begin{align}
    \mathrm{CCF}(\textit{v}) &= \sum_i \frac{1}{\tilde{s}^2}\left(\vec{d}_i - \tilde{\phi}\vec{m}_{i,\mathrm{w/o}\ X}\right)^T\Sigma_{0,i}^{-1}\ \left(\vec{m}_{i}-\vec{m}_{i,\mathrm{w/o}\ X}\right)(\textit{v}).
\end{align}
The cross-correlation functions (CCFs) were subsequently normalised to signal-to-noise units using samples at velocities outside of the expected peak ($|\textit{v}|>300\ \mathrm{km\ s^{-1}}$). 

\begin{figure}[h!]
    \centering
    \resizebox{\hsize}{!}{\includegraphics[width=17cm]{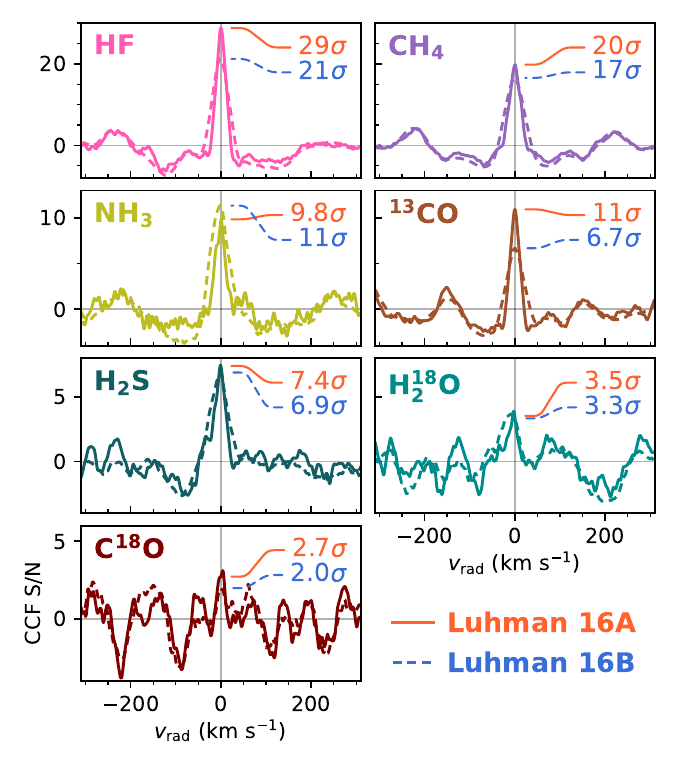}}
    \caption{Cross-correlation analysis of minor species in the spectra of Luhman 16A (solid) and B (dashed). The detection significance at $\textit{v}=0\ \mathrm{km\ s^{-1}}$ is indicated in the upper-right corner of each panel, with Luhman 16A at the top. The panel rows use different y-axis limits for legibility.}
    \label{fig:CCF}
\end{figure}

Figure~\ref{fig:CCF} shows the CCFs of the minor species. The calculations were performed in the brown dwarf rest frames, and thus we expect signals at $\textit{v}=0\ \mathrm{km\ s^{-1}}$. For both brown dwarfs, we find strong evidence of the presence of HF, CH$_4$, NH$_3$, $^{13}$CO, and H$_2$S, while CO$_2$, HCN, H$_2^{17}$O, and C$^{17}$O are not detected. The oxygen isotopologues H$_2^{18}$O and C$^{18}$O are measured at less conclusive significances of $\sim$\,$2$ to $3.5\sigma$. Hence, we compared the Bayesian information criteria (BICs) of the complete model with a retrieval where one of the species is excluded. The inclusion of H$_2^{18}$O and C$^{18}$O are not preferred for Luhman 16B with $\Delta\mathrm{BIC}=108.8$ and $253.1$, respectively. However, we find that the fit for the A component is significantly improved by H$_2^{18}$O ($\Delta\mathrm{BIC}=-41.2$), but not necessarily by C$^{18}$O ($\Delta\mathrm{BIC}=3.8$). We reached similar conclusions when comparing our results with the \texttt{MultiNest}-computed evidence, as well as with the Akaike and simplified Bayesian predictive information criteria (\citealt{Thorngren_ea_2025}). 

The estimated cross-correlation significances are affected by non-Gaussian statistics, which is apparent from the comparable out-of-peak signals for both brown dwarfs in Fig.~\ref{fig:CCF}. These systematic noise structures are produced by several effects. For strong contributors like CH$_4$ and $^{13}$CO, auto-correlation results in overestimated noise, thus leading to reduced detection significances. Furthermore, spectral residuals (see Figs.~\ref{fig:spectrum} and~\ref{fig:spectrum_app1}) from tellurics, poorly fitted lines of other molecules, or line-list deficiencies produce pseudo-stochastic variations in the CCFs that can mimic true signals. In this context, we evaluated the H$_2^{18}$O and C$^{18}$O significances using a bootstrap method, as described in Appendix~\ref{app:bootstrap_CCF}. We find that the weak signals are consistent with chance alignments between residuals and the molecular templates. For that reason, and in view of the discrepant Bayesian evidence results, we caution that the constrained H$_2^{18}$O and C$^{18}$O abundances are likely biased. 

\subsection{Elemental and isotopic abundances} \label{sect:elemental_abundances}
\begin{figure}[h!]
    \centering
    \resizebox{\hsize}{!}{\includegraphics[width=17cm]{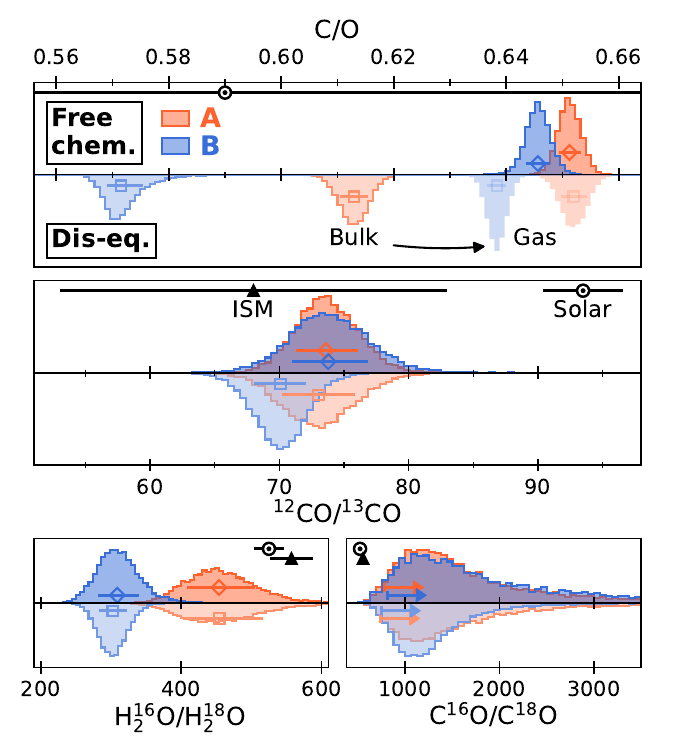}}
    \caption{Chemical and isotopic abundance ratios retrieved for Luhman 16AB. Both the results from the free-chemistry and disequilibrium retrievals are shown. For context, we show the abundance ratios of the local ISM \citep{Milam_ea_2005,Wilson_1999} and the Sun \citep{Asplund_ea_2021,Lyons_ea_2018}.}
    \label{fig:chem_ratios}
\end{figure}

The detection of multiple chemical species allowed us to constrain the elemental and isotopic abundances. Figure~\ref{fig:chem_ratios} presents the $\mathrm{C/O}$, $\mathrm{^{12}C/^{13}C}$, and $\mathrm{^{16}O/^{18}O}$ ratios of both targets. The free-chemistry retrieval constrains $\mathrm{C/O_A}=0.651^{+0.002}_{-0.002}$ and $\mathrm{C/O_B}=0.646^{+0.002}_{-0.002}$, which measure the gaseous content of the atmosphere and thus omit the oxygen that is condensed in silicate-oxide clouds. Both free-chemistry $\mathrm{C/O}$ ratios are therefore elevated compared to the disequilibrium retrievals seen in the inverted, lower panel of Fig.~\ref{fig:chem_ratios} ($\mathrm{C/O}_\text{A,Dis-eq.}=0.613^{+0.002}_{-0.002}$, $\mathrm{C/O}_\text{B,Dis-eq.}=0.572^{+0.004}_{-0.003}$). The disequilibrium model is defined by the bulk $\mathrm{C/O}$ ratio, which includes condensed oxygen and negligible condensed carbon. If only the gaseous species are considered, we recover similar gaseous ratios near the photosphere ($P\sim$\,$2\ \mathrm{bar}$; light-shaded posteriors in Fig.~\ref{fig:chem_ratios}), revealing oxygen sequestration of $6$--$10\%$. This falls somewhat short of the oxygen sink of $17.8^{+1.7}_{-2.3}\%$ predicted for the chemical composition of the solar neighbourhood \citet{Calamari_ea_2024}, but we note that reduced silicon abundances can still yield comparable levels of sequestration. The discrepant bulk ratios between the binary could result from inhomogeneous (gaseous) oxygen-rich regions that decrease the overall $\mathrm{C/O}$ ratio on the observed hemisphere of Luhman 16B. In this case, the uncertainties for the free-chemistry and disequilibrium models will underestimate the true dispersion in $\mathrm{C/O}$ ratios. An extended model with contribution from multiple columns (e.g. \citealt{Vos_ea_2023,de_Regt_ea_2025,Zhang_ea_2025}) may provide a solution, but is beyond the scope of this paper. 

The constrained carbon isotope ratios are remarkably similar between Luhman 16A and B, with $\mathrm{^{12}C/^{13}C_A}=74^{+2}_{-2}$ and $\mathrm{^{12}C/^{13}C_B}=74^{+3}_{-3}$, respectively. As a demonstration of its robustness, the carbon isotopologue ratio inferred with the free-chemistry model agrees within $2\sigma$ to that found in chemical disequilibrium. The oxygen isotopologue ratios are presented in the lower panels of Fig.~\ref{fig:chem_ratios}, where we find $\mathrm{H_2^{16}O/H_2^{18}O_A}=454^{+52}_{-45}$ and $\mathrm{H_2^{16}O/H_2^{18}O_B}=309^{+31}_{-27}$ from water, and lower limits of $\mathrm{C^{16}O/C^{18}O_A}>785$ and $\mathrm{C^{16}O/C^{18}O_B}>815$ from CO (2.5th percentile). As described in Sect.~\ref{sect:detections}, the H$_2^{18}$O abundance is likely biased due to poorly fitted H$_2^{16}$O lines, which explains the inconsistent H$_2$O and CO isotopologue ratios. 

\begin{figure}[h!]
    \centering
    \resizebox{\hsize}{!}{\includegraphics[width=17cm]{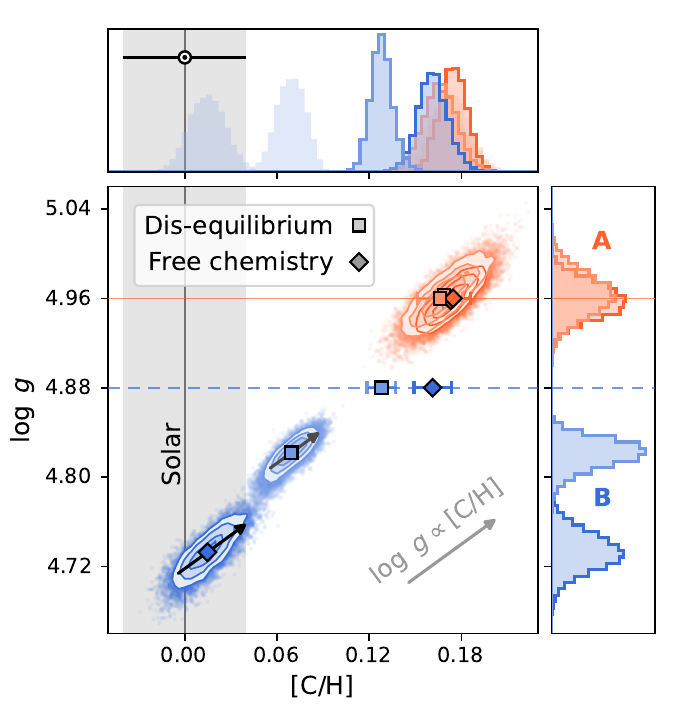}}
    \caption{Retrieved correlation between surface gravity and the carbon abundance, relative to hydrogen and the solar value (i.e. $\mathrm{[C/H]}$; \citealt{Asplund_ea_2021}). To a first order, these two parameters follow a linear relation and can be projected onto the expected $\log\textit{g}_\mathrm{A}=4.96$ and $\log\textit{g}_\mathrm{B}=4.88$.}
    \label{fig:logg_metallicity}
\end{figure}

We find strong correlations between the retrieved absolute abundances and surface gravities (see Fig.~\ref{fig:logg_metallicity}), which is often seen in atmospheric retrievals \citep{Zhang_ea_2021,de_Regt_ea_2024,Gonzalez_Picos_ea_2024,Zhang_ea_2024,Grasser_ea_2025} and stems from their inverse effect in calculating the optical depth ($\tau=\kappa P/\textit{g}$ with $\kappa\propto\mathrm{VMR}$; \citealt{Molliere_ea_2015}). In fact, for Luhman 16B the surface gravities constrained via the disequilibrium and free-chemistry retrievals ($\log\textit{g}_B=4.73^{+0.02}_{-0.02}$, $4.82^{+0.01}_{-0.01}$) are offset from each other and below the mean of the Gaussian prior ($\log\textit{g}_B=4.88\pm0.09$), which would hinder a comparison of the VMRs and metallicity. From a fixed-$\log\textit{g}$ test retrieval, we find that shallow slopes in the third and sixth spectral orders may bias the fit towards lower surface gravities. These slopes might arise from a more heterogeneous atmosphere on Luhman 16B compared to A, but we caution that high-resolution spectra are less sensitive to continuum shapes, which makes an atmospheric origin ambiguous. To resolve the issue of offset constraints, we linearly projected the abundance posteriors onto the expected surface gravities, $\log\textit{g}_\mathrm{A}=4.96$ and $\log\textit{g}_\mathrm{B}=4.88$ (see also \citealt{Gonzalez_Picos_ea_2025c}). Figure~\ref{fig:logg_metallicity} demonstrates how this projection brings the carbon abundance (relative to hydrogen and the solar value; $\mathrm{[C/H]}$) of the free-chemistry model into agreement between the two brown dwarfs. We note that the posterior samples in Fig.~\ref{fig:logg_metallicity} follow a somewhat steeper relation than $\log \textit{g}\propto\mathrm{[C/H]}$, likely due to additional degeneracies with the atmospheric temperature gradients. However, for the sake of consistency between elements and chemistry models, we used the linear relation to project the elemental abundances onto the expected surface gravities. As with carbon, the four other elements are brought into better agreement as a consequence of the $\log\textit{g}\propto\mathrm{[X/H]}$ projection (see Table \ref{tab:params}).

\begin{figure}[h!]
    \centering
    \resizebox{\hsize}{!}{\includegraphics[width=17cm]{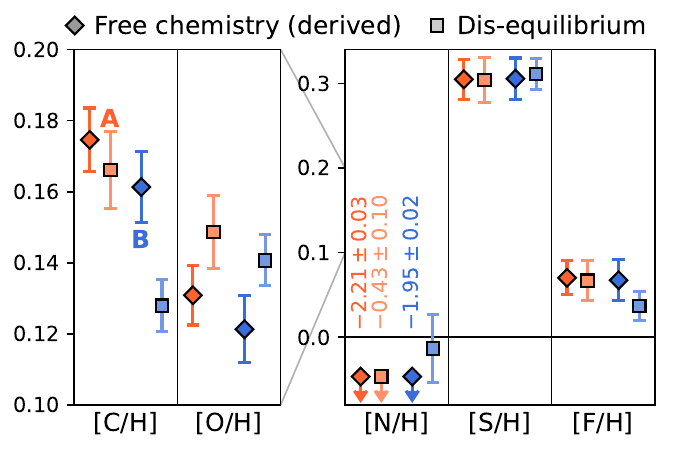}}
    \caption{Abundances of carbon, oxygen, nitrogen, sulphur, and fluorine as inferred from the detected gaseous molecules (see Sect.~\ref{sect:detections}). The values are shown relative to the solar composition and uncertainties reported by \citet{Asplund_ea_2021}.}
    \label{fig:abs_abund}
\end{figure}

Figure~\ref{fig:abs_abund} presents the elemental abundances as projected onto the expected surface gravities. The free-chemistry abundances are solely derived from the detected molecules, while the disequilibrium model represents the bulk composition. The free-chemistry model therefore greatly underestimates the nitrogen abundance due to missing measurements of the primary carrier, N$_2$, at the temperatures of L-T dwarfs (see Sect.~\ref{sect:gas_reservoir}). Interestingly, the concentrations of sulphur and fluorine are found to be largely consistent between chemistry models and across the binary, but they show different enrichments compared to carbon and oxygen. Except for nitrogen, both chemical models reveal elevated abundances compared to the solar composition, which therefore suggests that Luhman 16AB has a modest super-solar metallicity.  Using $\mathrm{[C/H]}$ as a metallicity proxy (see Sect.~\ref{sect:gas_reservoir}) results in a value of $\sim$\,$0.15,$ but we caution that this can be biased by the assumed surface gravities, in particular via the estimated radii ($1\ R_\mathrm{Jup}$), which are less certain than the dynamical masses. 

\subsection{Disequilibrium chemistry}
\begin{figure*}[h!]
    \centering
    \includegraphics[width=17cm]{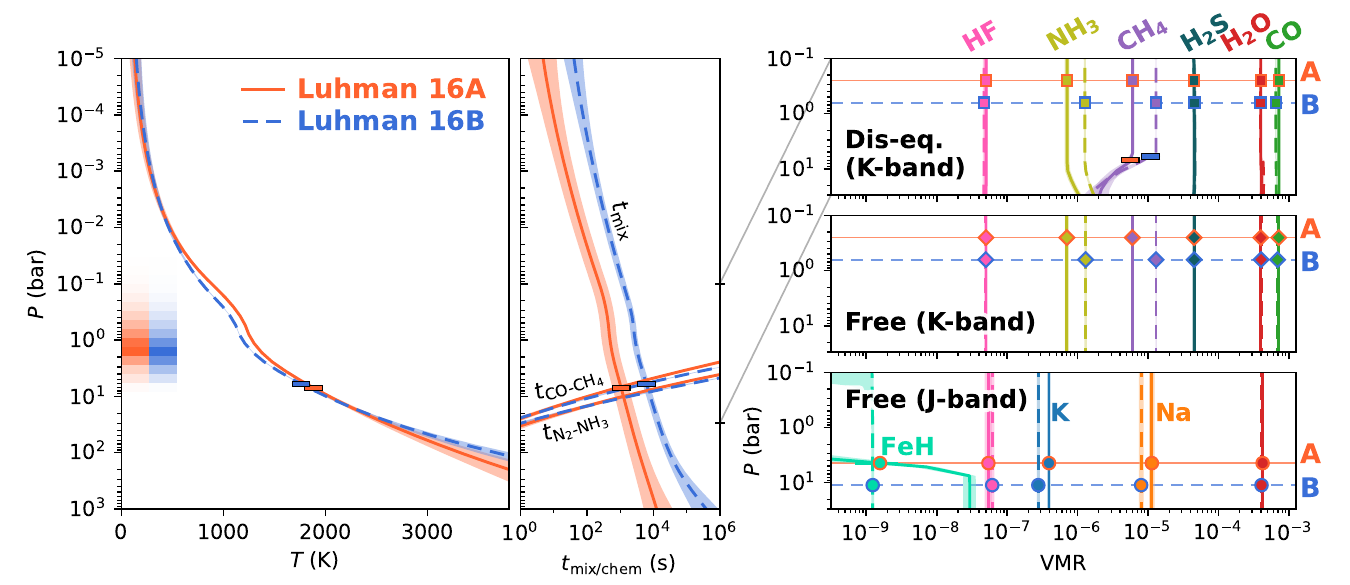}
    \caption{Comparison of temperature and abundance constraints. \textit{Left panel}: Temperature profiles of Luhman 16A and B, retrieved with the chemical disequilibrium model described in Sect.~\ref{sect:meth_chem}. Luhman 16A displays a heating of $\sim$\,$100\ \mathrm{K}$ near the photosphere, which is indicated by the shaded regions in the left side of the panel. \textit{Middle panel}: Mixing and chemical reaction timescales as functions of pressure. The intersection where vertical mixing becomes more efficient than the CO-CH$_4$ reaction is indicated by a horizontal bar. The higher $K_\mathrm{zz}$ for Luhman 16A is visible from its reduced mixing timescale. \textit{Right panels}: Cutouts of the chemical abundance profiles between $0.06$ and $30\ \mathrm{bar}$. \textit{From top to bottom}: Chemical disequilibrium and free-chemistry abundances from the K-band analysis of this work, and the CRIRES$^+$ J-band constraints presented in \citet{de_Regt_ea_2025}. The envelopes and error bars in all panels show the $68\%$ credible region.}
    \label{fig:eqchem_VMRs}
\end{figure*}

At the L-T transition, brown dwarfs can experience vigorous vertical mixing, which upsets their chemical equilibrium (e.g. \citealt{Hubeny_ea_2007,Zahnle_ea_2014,Mukherjee_ea_2022}). In this section we investigate the severity of disequilibrium in the Luhman 16 atmospheres by employing the \texttt{FastChem} model outlined in Sect.~\ref{sect:meth_chem}. Figure~\ref{fig:eqchem_VMRs} presents the retrieved temperature and abundance profiles, which are corrected for the surface-gravity degeneracy following Sect.~\ref{sect:elemental_abundances}. The right panels compare the disequilibrium abundances with the vertically constant free-chemistry abundances and shows good agreement between all molecules at the photospheres. 

Although most species show similar concentrations across the binary, CH$_4$ and NH$_3$ are distinctly more abundant in Luhman 16B. For L-T transition objects, CH$_4$ is particularly sensitive to the atmospheric dynamics as stronger vertical mixing leads to deeper quenching and thus a lower abundance \citep{Zahnle_ea_2014,Moses_ea_2016}. We find somewhat stronger mixing in Luhman 16A, which reduces the mixing timescale and causes a deeper quenching of CH$_4$ (and NH$_3$) as is illustrated by the middle panel of Fig.~\ref{fig:eqchem_VMRs}. The diffusivities -- which affect only the chemistry -- are constrained at $\log(K_\mathrm{zz}[\mathrm{cm^2\ s^{-1}}])=8.70^{+0.25}_{-0.21}$ and $8.17^{+0.12}_{-0.09}$ for Luhman 16A and B, respectively. This corresponds to quenching at $P_\mathrm{CO-CH_4-H_2O}=7.1^{+1.0}_{-0.8}$ and $6.0^{+0.3}_{-0.2}\ \mathrm{bar}$. The Luhman 16B atmosphere is cooler by $\sim$\,$100\ \mathrm{K}$ at these quench points, as evidenced by the left panel of Fig.~\ref{fig:eqchem_VMRs}. We hypothesise that the elevated temperature on the primary component may stem from the radiative feedback of more abundant clouds (e.g. \citealt{Morley_ea_2024}). The high mixing strengths in Luhman 16AB align well with the retrieval results for two older L-T transition objects, HD 4747B and DENIS J0255 \citep{Xuan_ea_2022,de_Regt_ea_2024}.

\section{Discussion} \label{sect:discussion}
\subsection{Chemistry across wavelengths}
The right panels of Fig.~\ref{fig:eqchem_VMRs} compare the chemical abundance constraints of this work with the measurements of \citet{de_Regt_ea_2025}, obtained from CRIRES$^+$ J-band spectra. Since the opacity cross-sections depend strongly on the probed wavelength range, we do not cover all molecules and atoms in both bands. Yet, we detect H$_2$O and HF at both wavelengths, for both brown dwarfs, and find equal mixing ratios. As the J band probes deeper into the atmosphere than the K band ($\sim$\,$10$ vs $1\ \mathrm{bar}$) the stable H$_2$O and HF abundances provide evidence of vertical homogenisation, which is also visible from the chemical disequilibrium profiles in the upper right panel of Fig.~\ref{fig:eqchem_VMRs}. 

Due to their close proximity to the Earth, Luhman 16A and B are the most studied brown dwarfs with observations that cover a wide range of wavelengths and spectral resolutions. Here, we put our CRIRES$^+$ J- and K-band analyses into context by reviewing the chemical constituents of the Luhman 16 atmospheres. Spectroscopic studies at optical wavelengths commonly find atomic lines from the alkalis Na, K, Rb, and Cs \citep{Luhman_2013,Heinze_ea_2021} as well as Li, which affirms the sub-stellar nature of the binary \citep{Faherty_ea_2014,Lodieu_ea_2015}. The bluer near-IR bands (Y, J) still probe Na and K lines, while absorption from the metal hydrides CrH and FeH is also identified in both brown dwarfs \citep{Burgasser_ea_2013,Burgasser_ea_2014,Faherty_ea_2014,Buenzli_ea_2015a,Buenzli_ea_2015b,Lodieu_ea_2015,Kellogg_ea_2017}. In \citet{de_Regt_ea_2025}, we present the first abundance constraints of FeH, Na and K (see Fig.~\ref{fig:eqchem_VMRs}). Despite the FeH detections at shorter wavelengths, the high-resolution H-band spectra of \citet{Ishikawa_ea_2025} do not show the strong FeH band head expected at $1582\ \mathrm{nm}$. Interestingly, this discrepancy may be explained by the rainout of gaseous FeH into iron clouds, which leads to a rapidly decreasing abundance with altitude \citep{Visscher_ea_2010,Burningham_ea_2021,Rowland_ea_2023}. Shorter wavelengths generally probe deeper in the atmosphere (e.g. \citealt{Vos_ea_2023,McCarthy_ea_2025}) where the higher temperatures prevent the complete condensation of FeH, while the higher H-band altitudes might be too depleted in FeH to be detectable. 

The volatile molecules H$_2$O, CO and CH$_4$ dominate the near-IR spectrum (e.g. \citealt{Crossfield_ea_2014,Biller_ea_2024,Chen_ea_2024,Chen_ea_2025}). While most spectral features are comparable between Luhman 16A and B, CH$_4$ absorption is notably stronger in the T-type component (e.g. \citealt{Burgasser_ea_2013,Faherty_ea_2014,Lodieu_ea_2015,Kellogg_ea_2017}). In this work, we confirm that CH$_4$ is indeed more abundant in the Luhman 16B atmosphere as a result of less efficient vertical mixing. The lower quench pressure also enhances the NH$_3$ abundance, which leads to somewhat deeper absorption in the K band, as noted by \citet{Ishikawa_ea_2025}. We highlight that the latest high-resolution spectra ($\mathcal{R}>10^4$) presented in \citet{Ishikawa_ea_2025}, \citet{de_Regt_ea_2025}, and this work have expanded the chemical inventory to include weakly contributing volatiles like NH$_3$, H$_2$S, HF, and $^{13}$CO. \citet{Ishikawa_ea_2025} also make the first direct measurement of the main atmospheric constituent, H$_2$, using lines at $2121.8$, $2223.3$, and $2406.6\ \mathrm{nm}$. We cannot reproduce this because the CRIRES$^+$ K2166 setting does not cover these wavelengths. Lastly, recent JWST variability monitoring has expanded the spectroscopic coverage up to $14\ \mathrm{\mu m}$ \citep{Biller_ea_2024,Chen_ea_2025}. These JWST spectra access more absorption bands of H$_2$O, CO, and CH$_4$ as well as the $8$--$11\ \mathrm{\mu m}$ silicate-oxide cloud feature that is weak for Luhman 16A and not detected for B, consistent with the expected clearing of photospheric condensates at the L-T transition \citep{Chen_ea_2025}.

\subsection{Implications for the formation history}
From the disequilibrium retrievals, we constrained bulk ratios of $\mathrm{C/O_A}=0.613^{+0.002}_{-0.002}$ and $\mathrm{C/O_B}=0.572^{+0.004}_{-0.003}$, which are consistent with the solar value ($0.59\pm0.08$; \citealt{Asplund_ea_2021}). To our knowledge, only one other member of the Oceanus moving group (bona fide or high-likelihood; \citealt{Gagne_ea_2023}), the G5 star HD 77006A, has a reported $\mathrm{C/O}$ of $0.51\pm0.03$ \citep{Takeda_2023}. This comparison suggests a possible tension between objects that formed in the same environment, but further $\mathrm{C/O}$ measurements of Oceanus members are needed to draw such conclusions. Moreover, we note that the inferred bulk $\mathrm{C/O}$ ratios depend on the condensation efficiency, for instance through the availability of magnesium and silicon \citep{Calamari_ea_2024} for which we assumed solar abundances in this work. The metallicities in the Oceanus group cluster near the solar value (e.g. $\mathrm{[Fe/H]}\sim$\,$-0.01$; \citealt{Takeda_2023}, $-0.03\pm0.03$, $0.10\pm0.09$, $0.08\pm0.05$; \citealt{Soubiran_ea_2022}, $0.10\pm0.18$; \citealt{Hejazi_ea_2022}, $\sim$\,$-0.50$; \citealt{Steinmetz_ea_2020}) but there is enough spread that our estimate of $\mathrm{[C/H]}\sim$\,$0.15$ for Luhman 16AB is congruent with the previous studies.

The similar $\mathrm{^{12}C/^{13}C}$ ratios that we find between the binary reinforce the presumed shared formation and evolution history. In addition to the biases discussed in Sect.~\ref{sect:detections}, the co-evolving context gives another reason to question the validity of the discrepant $\mathrm{H_2^{16}O}/\mathrm{H_2^{18}O}$ ratios that are retrieved between A and B. The carbon isotope ratios of Luhman 16AB ($\mathrm{^{12}C/^{13}C}\sim$\,$74$) are in line with the present-day, local ISM ($\mathrm{^{12}C/^{13}C_{ISM}}=68\pm15$; \citealt{Milam_ea_2005}) but lower than the solar wind ($\mathrm{^{12}C/^{13}C_\odot}=93.5\pm3.1$; \citealt{Lyons_ea_2018}). Models of galactic chemical evolution predict an enhancement of the interstellar $\mathrm{^{13}C}$ over time \citep{Romano_ea_2022} and recent observations of M-dwarf stars confirm this trend \citep{Gonzalez_Picos_ea_2025b}. The elevated $\mathrm{^{13}C}$ abundance we find for Luhman 16AB is thus compatible with recent inheritance from the ISM, which provides further evidence for the modestly young age of Luhman 16AB ($\sim$\,$500\ \mathrm{Myr}$; \citealt{Gagne_ea_2023}). 

Since the oxygen isotope is not confidently detected in the presented analysis, we refrain from interpreting the $\mathrm{^{16}O/^{18}O}$ ratios in the context of the Luhman 16AB formation history. The high rotational velocities of Luhman 16AB ($\textit{v}\sin i=14.76\pm0.03$ and $24.79\pm0.04\ \mathrm{km\ s^{-1}}$) hinder the identification of C$^{18}$O and H$_2^{18}$O absorption lines as they become blended with neighbouring features. Slower rotators typically present tentative to more credible evidence of minor isotopologues, even in K-band spectra of lower quality compared to those of Luhman 16AB \citep{Zhang_ea_2022,Xuan_ea_2024,Gonzalez_Picos_ea_2024,Gonzalez_Picos_ea_2025b,Mulder_ea_2025,Grasser_ea_2025}. The strongest CO absorption is found at its fundamental band ($4.3$--$5.1\ \mathrm{\mu m}$) and can be accessed with high-resolution M-band spectroscopy \citep{Crossfield_ea_2019}. Though Luhman 16A and B are bright enough, ground-based observations suffer from the high thermal sky background \citep{Parker_ea_2024}. Alternatively, recent medium-resolution JWST spectra ($\mathcal{R}\sim$\,$3\,000$) have uncovered several minor isotopologues, including $^{13}$CO, C$^{18}$O, C$^{17}$O \citep{Gandhi_ea_2023,Lew_ea_2024,Molliere_ea_2025,Gonzalez_Picos_ea_2025c}, $^{15}$NH$_3$ \citep{Barrado_ea_2023,Kuhnle_ea_2025}, as well as CH$_3$D \citep{Rowland_ea_2024}. Given the high brightness, it is reasonable to expect that these isotopologues are also detectable in the Luhman 16AB atmospheres. 

\begin{figure*}[h!]
    \centering
    \includegraphics[width=17cm]{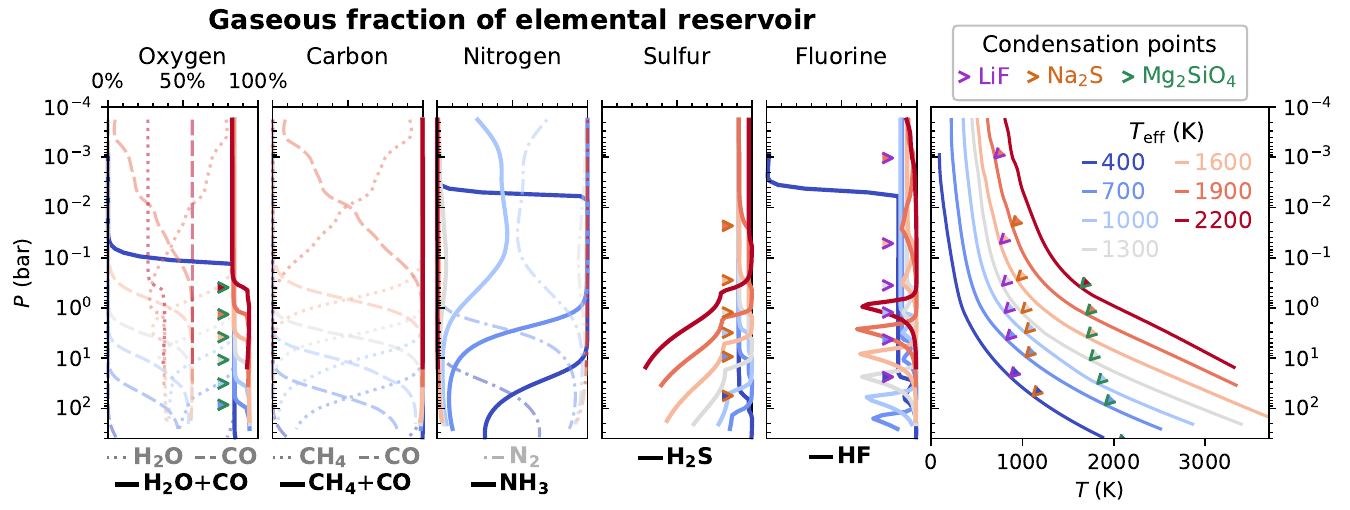}
    \caption{Abundance profiles of gases as a fraction of the total elemental reservoir of oxygen, carbon, nitrogen, sulphur, and fluorine. The abundances are calculated with \texttt{FastChem Cond} \citep{Kitzmann_ea_2024}, using SONORA Elf Owl temperature structures (shown in the right panel; \citealt{Mukherjee_ea_2024}) that are representative of Y-, T-, L- and late M-type dwarfs. Oxygen and carbon are carried by multiple gaseous molecules (H$_2$O, CO, and CH$_4$), which are shown as their separate and summed profiles. The condensation points of important cloud species are indicated as arrowheads for reference.}
    \label{fig:gas_fraction}
\end{figure*}

\subsection{Detectable reservoir of elements} \label{sect:gas_reservoir}
Measurements of molecular gases in sub-stellar atmospheres do not always reflect the bulk elemental composition because the chemistry might favour the formation of other, unobserved gases or condensates. Here, we investigate which molecules reliably trace the elemental reservoirs across late M-, L-, T-, and Y-type temperature profiles, obtained from the SONORA Elf Owl modelling suite \citep{Mukherjee_ea_2024}. We employed \texttt{FastChem} \citep{Kitzmann_ea_2024} to compute the chemical equilibrium and include the effects of condensation and rainout. The input consists of a proto-solar composition, as tabulated by \citet{Bergemann_ea_2025}. Combined with the output number densities, we calculated the fraction of oxygen, carbon, nitrogen, sulphur, and fluorine that is carried by selected molecular gases, and this is presented in Fig.~\ref{fig:gas_fraction}. As an example, in the hotter ($\leq$\,T1) atmospheres NH$_3$ probes $\lesssim$\,$10\%$ of the nitrogen because this element primarily exists in undetectable N$_2$ gas \citep{Lodders_ea_2002}. This implies that NH$_3$ will underestimate the total nitrogen concentration, as we found from the free-chemistry model in Sect.~\ref{sect:elemental_abundances}. Similarly for higher temperatures, sulphur converts to SH and atomic S gas \citep{Visscher_ea_2006}, leading to a decreasing H$_2$S fraction with depth. However, at the Luhman 16AB temperatures H$_2$S is the predominant sulphur carrier, which explains the consistent $\mathrm{[S/H]}$ derived for the free-chemistry model and its retrieved value in the equilibrium model, as noted in Sect.~\ref{sect:elemental_abundances}.

The oxygen sequestration outlined in Sect.~\ref{sect:elemental_abundances} is demonstrated by the gas reduction that is seen at the Mg$_2$SiO$_4$ condensation points (green arrows) in Fig.~\ref{fig:gas_fraction}. Above those points, the summed H$_2$O and CO measurements can account for $\sim$\,$80\%$ of the oxygen concentration in the calculated models, but we note that the extent of oxygen-trapping depends on assumptions of the condensation physics such as the availability of magnesium and silicon \citep{Calamari_ea_2024}. Below the condensation points, gaseous SiO holds $\sim$\,$6\%$ of the oxygen budget. The Y-dwarf ($T_\mathrm{eff}=$\,$400\ \mathrm{K}$) reveals eventual depletions of nitrogen and oxygen when ammonia and water ices form, respectively \citep{Morley_ea_2014}. Fortunately, the carbon reservoir is measured fully by summing the CH$_4$ and CO abundances. For this reason, we recommend the adoption of $\mathrm{[C/H]}$ as a proxy for the metallicity when both CO and CH$_4$ can be measured. It should be noted that the \texttt{FastChem} models shown in Fig.~\ref{fig:gas_fraction} do not consider disequilibrium chemistry although that should minimally affect the summed abundances of $\mathrm{H_2O}+\mathrm{CO}$ and $\mathrm{CH_4}+\mathrm{CO}$ or the general trends observed. While CO$_2$ holds a negligible fraction of carbon and oxygen at the assumed protosolar composition ($<0.1\%$), we observe that its contribution increases up to $\sim$\,$8\%$ for both elements at $100$ times the solar metallicity.

Hydrogen-fluoride (HF) has recently been detected in the high-resolution spectra of several M dwarfs \citep{Gonzalez_Picos_ea_2025b}, brown dwarfs \citep{Gonzalez_Picos_ea_2024,Mulder_ea_2025,Grasser_ea_2025}, and high-mass exoplanets \citep{Gonzalez_Picos_ea_2025a,Zhang_ea_2024}. Notably, the molecule appears detectable for a broad range of spectral types, i.e. M0--T0.5. Luhman 16A and B are the first sub-stellar objects where HF is detected at J- and K-band wavelengths \citep{de_Regt_ea_2025,Ishikawa_ea_2025}. This observed stability with temperature and pressure can be explained using Fig.~\ref{fig:gas_fraction}. For each temperature profile, we find that fluorine primarily exists in the form of gaseous HF. The production of AlF gas temporarily reduces the HF fraction to $\sim$\,$65\%$, but the subsequent rainout of Al-bearing condensates releases fluorine back to form its hydride. Only the Y-dwarf shows a depletion of HF into gaseous H$_6$F$_6$ at $\lesssim$\,$10^{-2}\ \mathrm{bar}$. Observations of the cold ISM also find that HF can account for most of the gas-phase fluorine \citep{Gerin_ea_2016}. In view of the stability with cooler temperatures, we predict that HF remains detectable in high-resolution J-band spectra of late T-dwarfs (at $\gtrsim$\,$1.25\ \mathrm{\mu m}$). In the K band, however, the strong CH$_4$ absorption (e.g. \citealt{Cushing_ea_2005}) likely masks the HF lines beyond $\gtrsim$\,$2.3\ \mathrm{\mu m}$. 

Previous studies have suggested that fluorine can bond with lithium to form a considerable amount of LiF, both as a gas \citet{Gharib-Nezhad_ea_2021} and a condensate species \citep{Lodders_ea_1999,Lodders_ea_2006}. The creation of LiF, in addition to other lithium compounds (LiCl, LiOH, and LiH; \citealt{Gharib-Nezhad_ea_2021}), could thus contribute to the observed weakening of the optical Li doublet for T-type dwarfs \citep{Kirkpatrick_ea_2008,Martin_ea_2022}. The presented \texttt{FastChem} equilibrium models account for the gas- and condensed phase chemistry of lithium as implemented in Kitzmann et al. (\textit{in prep.}). Consequently, Fig.~\ref{fig:gas_fraction} shows small depletions of HF ($\sim$\,$10\%$) due to the formation of LiF clouds. Lithium is entirely condensed above these points because it is the rarer element compared to fluorine in the proto-solar composition ($A(\mathrm{Li})=3.39\pm0.02$, $A(\mathrm{F})=4.75\pm0.09$; \citealt{Bergemann_ea_2025}). The expected reduction in the HF mixing ratio is only $\sim$\,$0.03\ \mathrm{dex,}$ which is outside of most detection limits, even for the nearest brown dwarfs (see Table~\ref{tab:params}). For that reason, it appears that gaseous HF measurements cannot provide meaningful constraints on the depletion of lithium. Since most of the fluorine exists as HF, however, this molecule can still serve as a general tracer of metal enrichment. 

\section{Conclusions} \label{sect:conclusions}
We analysed high-resolution K-band spectra to infer the chemical properties of the Luhman 16AB atmospheres. The CRIRES$^+$ observations resolve the binary brown dwarfs and can be fitted extremely well using our retrieval framework. The fitted models result in detections of H$_2$O, $^{12}$CO, CH$_4$, H$_2$S, NH$_3$, HF, and the $^{13}$CO isotopologue. The molecular abundances retrieved using a free-chemistry and disequilibrium modelling approach are consistent with one another. The combined CO and CH$_4$ concentrations reveal an unmistakable chemical disequilibrium in the Luhman 16AB atmospheres, while the stronger CH$_4$ and NH$_3$ absorption of component B can be attributed to less vigorous mixing ($K_\mathrm{zz,A}\sim$\,$10^{8.7}$ vs $K_\mathrm{zz,B}\sim$\,$10^{8.2}\ \mathrm{cm^2\ s^{-1}}$). We reviewed the recent detections of HF in sub-stellar atmospheres using chemical equilibrium models. Most fluorine exists in the form of gaseous HF, and this molecule is stable with temperature and pressure, thereby explaining the detections for M-, L-, and T-type dwarfs in addition to the constant abundances that we infer at different depths of Luhman 16AB. 

The abundance constraints allowed us to derive bulk carbon-to-oxygen ratios of $\mathrm{C/O_A}=0.613^{+0.002}_{-0.002}$ and $\mathrm{C/O_B}=0.572^{+0.004}_{-0.003}$, in line with the solar composition. The retrieved gaseous $\mathrm{C/O}$ ratios are higher since oxygen is partially trapped in the silicate-oxide clouds. The $^{13}$CO detection reveals isotope ratios of $\mathrm{^{12}C/^{13}C_A}=74^{+2}_{-2}$ and $\mathrm{^{12}C/^{13}C_B}=74^{+3}_{-3}$, which are enriched in $^{13}$C compared to the Sun. The coincident isotope ratios provide further support for the idea that Luhman 16A and B have the same formation histories. Since evolved stars are predicted to enhance the interstellar $^{13}$C over time, and $\mathrm{^{12}C/^{13}C}\sim$\,$74$ aligns with the present-day ISM, the inferred ratios are compatible with a recent inheritance and therefore corroborate the moderately young age of Luhman 16AB ($\sim$\,$500\ \mathrm{Myr}$; \citealt{Gagne_ea_2023}). In line with a recent formation, the elemental abundances of carbon, oxygen, sulphur, and fluorine reveal an overall enrichment of metals compared to the solar composition. The metallicity is measured at $\mathrm{[C/H]}\sim$\,$0.15$, but we caution that this value is degenerate with the assumed surface gravity.

This analysis of Luhman 16AB presents some of the best elemental and isotopic abundance constraints of the ESO SupJup Survey and, broadly, for brown dwarfs. It remains challenging, however, to comprehensively understand the atmospheric chemistries. For instance, future studies could explore the chemical abundances in three (or four) dimensions. At the L-T transition in particular, the coupling between clouds and the temperature profile might result in heterogeneous mixing ratios with altitude, longitude, and latitude \citep{Lee_ea_2024, Teinturier_ea_2025}. As the closest known brown dwarfs, Luhman 16A and B are prime targets for such in-depth analyses using both high-resolution and broad-wavelength spectroscopy.

\begin{acknowledgements}
We thank the anonymous referee for their constructive feedback.
We thank Daniel Kitzmann for sharing the latest input files and his help in including lithium in \texttt{FastChem}. 
We also thank Jackie Faherty and Johanna Vos for a helpful discussion that motivated this addition.
S.d.R. and I.S. acknowledge NWO grant OCENW.M.21.010. 
Based on observations collected at the European Organisation for Astronomical Research in the Southern Hemisphere under ESO programme(s) 110.23RW.002.
This work used the Dutch national e-infrastructure with the support of the SURF Cooperative using grant no. EINF-4556 and EINF-9460. 
This research has made use of the Astrophysics Data System, funded by NASA under Cooperative Agreement 80NSSC21M00561.
\newline
\textit{Software}: Astropy \citep{Astropy_Collaboration_ea_2022}, corner \citep{Foreman_Mackey_2016}, FastChem \citep{Kitzmann_ea_2024}, Matplotlib \citep{Hunter_2007}, NumPy \citep{Harris_ea_2020}, petitRADTRANS \citep{Molliere_ea_2019}, PyMultiNest \citep{Feroz_ea_2009,Buchner_ea_2014}, SciPy \citep{Virtanen_ea_2020}.
\end{acknowledgements}

\bibliographystyle{aa}
\bibliography{references.bib}

\begin{appendix} 
\section{Cross-correlation bootstrap validation} \label{app:bootstrap_CCF}
\begin{figure}[h!]
    \centering
    \resizebox{\hsize}{!}{\includegraphics[width=17cm]{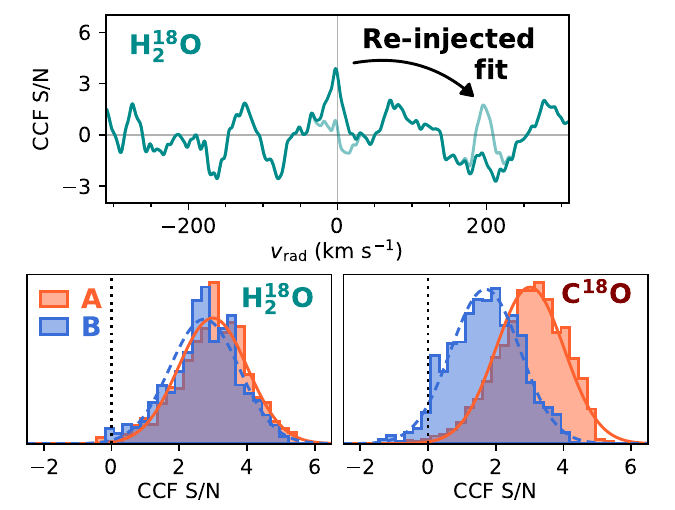}}
    \caption{Bootstrap validation of H$_2^{18}$O and C$^{18}$O cross-correlation signals. \textit{Top panel}: Example of the removal and re-injection of the H$_2^{18}$O CCF peak for Luhman 16A. \textit{Bottom panels}: Cross-correlation $\mathrm{S/N}$ distributions measured after injections between $\pm1000\ \mathrm{km\ s^{-1}}$.}
    \label{fig:bootstrap_CCF}
\end{figure}

As explained in Sect.~\ref{sect:detections}, several effects can result in the pseudo-stochastic CCF structure observed for both brown dwarfs in Fig.~\ref{fig:CCF}. Particularly for weak contributors, whose line depths are smaller than some spectral residuals, it is possible that random alignments can result in a $2$ to $4\sigma$ peak at the expected velocity. To evaluate the detectability of the signal at other velocities, we fitted Gaussian profiles to the H$_2^{18}$O and C$^{18}$O CCFs at $\textit{v}_\mathrm{rad}=0\ \mathrm{km\ s^{-1}}$. As shown in the left panel of Fig.~\ref{fig:bootstrap_CCF}, the fitted peak is subtracted and re-injected at another velocity, where the $\mathrm{S/N}$ is evaluated again. Following a bootstrap method (e.g. \citealt{Redfield_ea_2008,Hoeijmakers_ea_2020}), this injection and evaluation is performed between $-1000$ and $+1000\ \mathrm{km\ s^{-1}}$ in steps of $1\ \mathrm{km\ s^{-1}}$. The resulting $\mathrm{S/N}$ distributions of Luhman 16A and B are presented in Fig.~\ref{fig:bootstrap_CCF}. Since the distributions are not clearly separated from a significance of $0\sigma$, we cannot confidently claim detections of these molecules.

\section{Reduced equilibrium-chemistry model} \label{app:fastchem_validation}
We created a reduced input for the equilibrium-chemistry code \texttt{FastChem Cond} \citep{Kitzmann_ea_2024} in an effort to shorten computation times to be suitable for usage during a retrieval. Table~\ref{tab:fastchem_reduced} shows the included atomic gases, molecular gases, and condensate species. These are chosen based on their prevalence or impact on the gaseous compositions of L- or early T-type brown-dwarf atmospheres. This reduction from 618 gases and 235 condensates in the \texttt{FastChem} extended input down to, respectively, 43 and 16 species accelerates the computation by at least $10$ times. Depending on the number of atmospheric layers, temperatures, and CPU clock speed, the reduced input with rainout condensation typically converges in $\lesssim$\,$0.2\ \mathrm{s}$. 

\begin{figure*}[h!]
    \centering
    \includegraphics[width=17cm]{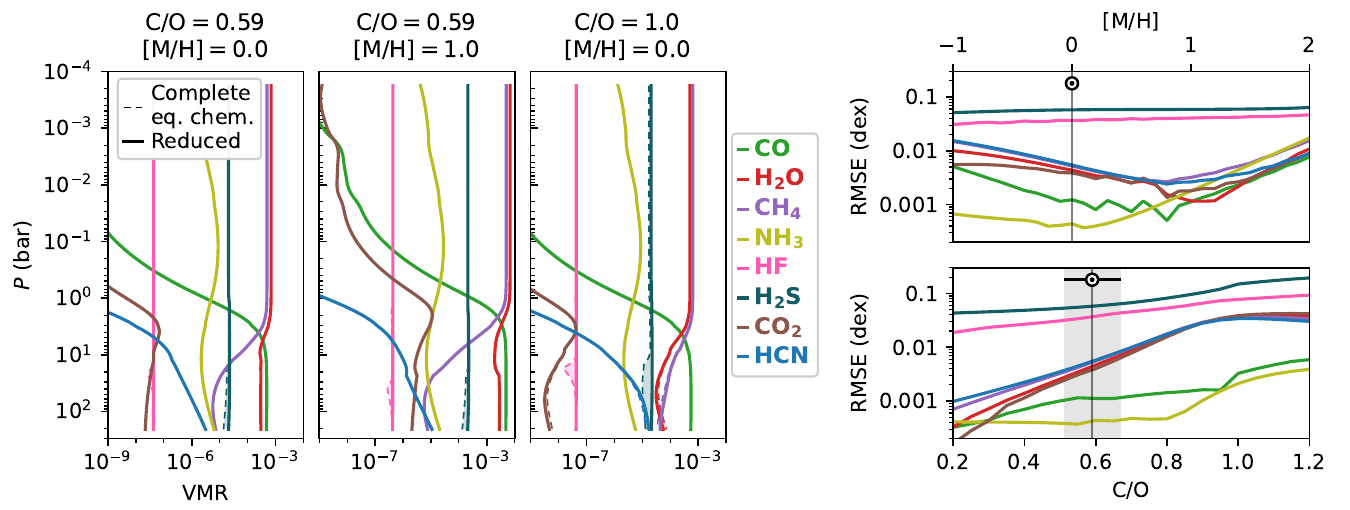}
    \caption{Accuracy validation of reduced \texttt{FastChem} input compared to the complete, extended equilibrium-chemistry model. \textit{Left panels}: Deviation in VMRs for a solar composition \citep{Asplund_ea_2021}, a $10$ times solar metallicity, and a $\mathrm{C/O}=1.0$, assuming a $T_\mathrm{eff}=1300\ \mathrm{K}$ temperature profile \citep{Mukherjee_ea_2024}. \textit{Right panels}: RMSE between the $\log \mathrm{VMRs}$ of the two inputs at varying metallicities and $\mathrm{C/O}$ ratios.}
    \label{fig:reduced_chemistry}
\end{figure*}

Naturally, the reduced model comes at a cost of accuracy compared to the extended \texttt{FastChem} input. Figure~\ref{fig:reduced_chemistry} presents the deviations between the reduced (solid) and complete (dashed) reaction networks. The abundance profiles of these molecules minimally differ between the two approaches, especially at $\lesssim$\,$10\ \mathrm{bar}$. We quantified the deviation in $\log \mathrm{VMRs}$ across all pressures via the root mean square error (RMSE), and present this for each molecule at varying metallicities and $\mathrm{C/O}$ ratios in the right panels of Fig.~\ref{fig:reduced_chemistry}. Apart from the deviation of H$_2$S and HF in the lower atmosphere, the tested compositions yield $\mathrm{RMSE}<$\,$0.05\ \mathrm{dex}$ for the other molecules. We therefore conclude that the reduced equilibrium-chemistry model is sufficiently accurate and can be employed for L-T transition objects with plausible compositions.

\begin{table}[h!]
    \small
    \centering
    \caption{Gaseous and condensate species included in our reduced \texttt{FastChem} input, sorted by mass.}
    \label{tab:fastchem_reduced}
    \begin{tabular}{r|l}
        \hline\hline
        Atomic gases & H, He, C, N, O, F, Na, Mg, Al, Si, S, Cl, \\
                     & K, Ca, Ti, V, Cr, Mn, Fe \\
        \hline
        Molecular gases & H$_2$, CH$_4$, OH, NH$_3$, H$_2$O, HF, NaH, \\
                        & C$_2$H$_2$, HCN, HNC, AlH, CO, N$_2$, C$_2$H$_4$, \\
                        & H$_2$S, HCl, CaH, AlO, CO$_2$, SiO, CrH, \\
                        & FeH, TiO, VO \\
        \hline
        Condensates & Cr(s,l), Fe(s,l), MgO(s,l), SiO(s), \\
                    & NaCl(s,l), VO(s,l), KCl(s,l), Na$_2$S(s,l), \\
                    & TiO$_2$(s,l), MnS(s), MgSiO$_3$(s,l), \\
                    & Al$_2$O$_3$(s,l), K$_2$S(s,l), CaTiO$_3$(s), \\
                    & Mg$_2$SiO$_4$(s,l), Ca$_2$SiO$_4$(s) \\
        \hline
    \end{tabular}
\end{table}

\newpage
\onecolumn

\newcommand{\twoentries}[2]{\begin{tabular}{@{}r@{}}#1\\\big[{\footnotesize #2}\big]\end{tabular}}
\section{Extended retrieval results} \label{app:extended_results}
\begin{longtable}{lll|rr|rr}
    \caption{Retrieved parameters and their uncertainties.} \label{tab:params} \\
    
    \hline
    \hline
    \textbf{Parameter} & \textbf{Description} & \textbf{Prior} & \multicolumn{2}{c|}{\textbf{Luhman 16A}} & \multicolumn{2}{c}{\textbf{Luhman 16B}} \\
     & & & \textbf{Free chem.} & \textbf{Dis-eq.} & \textbf{Free chem.} & \textbf{Dis-eq.} \\
    \hline
    \endfirsthead

    \caption{Continued.} \\
    
    \hline
    \hline
    \textbf{Parameter} & \textbf{Description} & \textbf{Prior} & \multicolumn{2}{c|}{\textbf{Luhman 16A}} & \multicolumn{2}{c}{\textbf{Luhman 16B}} \\
     & & & \textbf{Free chem.} & \textbf{Dis-eq.} & \textbf{Free chem.} & \textbf{Dis-eq.} \\
    \hline
    \endhead

    \hline
    \endfoot
    
     & & & \underline{Retrieved} & \underline{Derived$^{\rm (a)}$} & \underline{Retrieved} & \underline{Derived$^{\rm (a)}$} \\
    $\log\mathrm{H_2O}$      & H$_2$O abundance & $\mathcal{U}(-14.0,-2.0)$ & $-3.402^{+0.012}_{-0.012}$ & $-3.397^{+0.016}_{-0.015}$ & $-3.541^{+0.012}_{-0.012}$ & $-3.463^{+0.009}_{-0.010}$ \\
    $\log\mathrm{H_2^{18}O}$ & H$_2^{18}$O abundance & $\mathcal{U}(-14.0,-2.0)$ & $-6.06^{+0.05}_{-0.05}$ & $-6.06^{+0.06}_{-0.06}$ & $-6.03^{+0.04}_{-0.04}$ & $-5.94^{+0.03}_{-0.03}$ \\
    $\log\mathrm{H_2^{17}O}$ & H$_2^{17}$O abundance & $\mathcal{U}(-14.0,-2.0)$ & $<-7.53$ & $<-7.01$ & $<-7.00$ & $<-6.80$ \\
    $\log\mathrm{^{12}CO}$   & $^{12}$CO abundance & $\mathcal{U}(-14.0,-2.0)$ & $-3.146^{+0.012}_{-0.012}$ & $-3.140^{+0.015}_{-0.014}$ & $-3.309^{+0.012}_{-0.012}$ & $-3.245^{+0.009}_{-0.010}$ \\
    $\log\mathrm{^{13}CO}$   & $^{13}$CO abundance & $\mathcal{U}(-14.0,-2.0)$ & $-5.01^{+0.02}_{-0.02}$ & $-5.01^{+0.02}_{-0.02}$ & $-5.18^{+0.02}_{-0.02}$ & $-5.10^{+0.02}_{-0.02}$ \\
    $\log\mathrm{C^{18}O}$   & C$^{18}$O abundance & $\mathcal{U}(-14.0,-2.0)$ & $<-6.04$ & $-6.30^{+0.16}_{-0.23}$ & $<-6.22$ & $-6.34^{+0.11}_{-0.14}$ \\
    $\log\mathrm{C^{17}O}$   & C$^{17}$O abundance & $\mathcal{U}(-14.0,-2.0)$ & $<-6.42$ & $<-6.31$ & $<-6.46$ & $<-6.36$ \\
    $\log\mathrm{CH_4}$      & CH$_4$ abundance & $\mathcal{U}(-14.0,-2.0)$ & $-5.22^{+0.02}_{-0.02}$ & $-5.22^{+0.02}_{-0.02}$ & $-5.03^{+0.01}_{-0.01}$ & $-4.95^{+0.01}_{-0.01}$ \\
    $\log\mathrm{NH_3}$      & NH$_3$ abundance & $\mathcal{U}(-14.0,-2.0)$ & $-6.15^{+0.03}_{-0.03}$ & $-6.15^{+0.04}_{-0.04}$ & $-6.03^{+0.02}_{-0.02}$ & $-5.95^{+0.01}_{-0.02}$ \\
    $\log\mathrm{H_2S}$      & H$_2$S abundance & $\mathcal{U}(-14.0,-2.0)$ & $-4.35^{+0.03}_{-0.03}$ & $-4.34^{+0.03}_{-0.03}$ & $-4.49^{+0.03}_{-0.03}$ & $-4.40^{+0.02}_{-0.02}$ \\
    $\log\mathrm{HF}$        & HF abundance & $\mathcal{U}(-14.0,-2.0)$ & $-7.30^{+0.02}_{-0.02}$ & $-7.30^{+0.03}_{-0.03}$ & $-7.45^{+0.02}_{-0.03}$ & $-7.39^{+0.02}_{-0.02}$ \\
    $\log\mathrm{CO_2}$      & CO$_2$ abundance & $\mathcal{U}(-14.0,-2.0)$ & $<-5.37$ & $-7.09^{+0.03}_{-0.03}$ & $-5.37^{+0.08}_{-0.09}$ & $-7.20^{+0.02}_{-0.02}$ \\
    $\log\mathrm{HCN}$       & HCN abundance & $\mathcal{U}(-14.0,-2.0)$ & $<-6.28$ & $-6.99^{+0.05}_{-0.05}$ & $<-6.52$ & $-6.97^{+0.02}_{-0.01}$ \\
     & & & \underline{\twoentries{Derived$^{\rm (b)}$}{Projected$^{\rm (c)}$}} & \underline{\twoentries{Retrieved}{Projected$^{\rm (c)}$}} & \underline{\twoentries{Derived$^{\rm (b)}$}{Projected$^{\rm (c)}$}} & \underline{\twoentries{Retrieved}{Projected$^{\rm (c)}$}} \\
    $\mathrm{[C/H]}$        & Carbon abundance & $\mathcal{U}(-1.0,2.0)$ & \twoentries{$0.173^{+0.012}_{-0.012}$}{$0.175^{+0.009}_{-0.009}$} & \twoentries{$0.169^{+0.015}_{-0.014}$}{$0.166^{+0.011}_{-0.011}$} & 
                              \twoentries{$0.015^{+0.012}_{-0.012}$}{$0.161^{+0.010}_{-0.009}$} & \twoentries{$0.069^{+0.009}_{-0.010}$}{$0.128^{+0.007}_{-0.007}$} \\
    $\mathrm{[O/H]}$        & Oxygen abundance & $\mathcal{U}(-1.0,2.0)$ & \twoentries{$0.129^{+0.012}_{-0.012}$}{$0.131^{+0.008}_{-0.008}$} & \twoentries{$0.151^{+0.015}_{-0.013}$}{$0.149^{+0.010}_{-0.010}$} & 
                              \twoentries{$-0.025^{+0.012}_{-0.012}$}{$0.121^{+0.009}_{-0.009}$} & \twoentries{$0.082^{+0.009}_{-0.010}$}{$0.141^{+0.007}_{-0.007}$} \\
    $\mathrm{[N/H]}$        & Nitrogen abundance & $\mathcal{U}(-1.0,2.0)$ & \twoentries{$-2.21^{+0.03}_{-0.03}$}{$-2.21^{+0.03}_{-0.03}$} & \twoentries{$-0.43^{+0.09}_{-0.10}$}{$-0.43^{+0.09}_{-0.10}$} & 
                              \twoentries{$-2.09^{+0.02}_{-0.02}$}{$-1.95^{+0.02}_{-0.02}$} & \twoentries{$-0.07^{+0.04}_{-0.04}$}{$-0.01^{+0.04}_{-0.04}$} \\
    $\mathrm{[S/H]}$        & Sulphur abundance & $\mathcal{U}(-1.0,2.0)$ & \twoentries{$0.30^{+0.03}_{-0.03}$}{$0.30^{+0.02}_{-0.02}$} & \twoentries{$0.31^{+0.03}_{-0.03}$}{$0.30^{+0.03}_{-0.03}$} & 
                              \twoentries{$0.16^{+0.03}_{-0.03}$}{$0.31^{+0.02}_{-0.02}$} & \twoentries{$0.25^{+0.02}_{-0.02}$}{$0.31^{+0.02}_{-0.02}$} \\
    $\mathrm{[F/H]}$        & Fluorine abundance & $\mathcal{U}(-1.0,2.0)$ & \twoentries{$0.07^{+0.02}_{-0.02}$}{$0.07^{+0.02}_{-0.02}$} & \twoentries{$0.07^{+0.03}_{-0.03}$}{$0.07^{+0.02}_{-0.02}$} & 
                              \twoentries{$-0.08^{+0.02}_{-0.03}$}{$0.07^{+0.02}_{-0.02}$} & \twoentries{$-0.02^{+0.02}_{-0.02}$}{$0.04^{+0.02}_{-0.02}$} \\
    $\log\mathrm{^{12}CO/^{13}CO}$ & $\mathrm{^{12}CO/^{13}CO}$ ratio & $\mathcal{U}(0.0,5.0)$ & $1.87^{+0.01}_{-0.01}$ & $1.86^{+0.02}_{-0.02}$ & $1.87^{+0.02}_{-0.02}$ & $1.85^{+0.01}_{-0.01}$ \\
    $\log\mathrm{C^{16}O/C^{18}O}$ & $\mathrm{C^{16}O/C^{18}O}$ ratio & $\mathcal{U}(0.0,5.0)$ & $>3.04$ & $3.15^{+0.23}_{-0.16}$ & $>3.08$ & $3.09^{+0.14}_{-0.11}$ \\
    $\log\mathrm{C^{16}O/C^{17}O}$ & $\mathrm{C^{16}O/C^{17}O}$ ratio & $\mathcal{U}(0.0,5.0)$ & $>3.86$ & $>3.35$ & $>3.56$ & $>3.26$ \\
    $\log\mathrm{H_2^{16}O/H_2^{18}O}$ & $\mathrm{H_2^{16}O/H_2^{18}O}$ ratio & $\mathcal{U}(0.0,5.0)$ & $2.66^{+0.05}_{-0.05}$ & $2.66^{+0.06}_{-0.05}$ & $2.49^{+0.04}_{-0.04}$ & $2.48^{+0.03}_{-0.03}$ \\
    $\log\mathrm{H_2^{16}O/H_2^{17}O}$ & $\mathrm{H_2^{16}O/H_2^{17}O}$ ratio & $\mathcal{U}(0.0,5.0)$ & $>5.37$ & $>3.97$ & $>4.58$ & $>3.63$ \\
    $\log K_\mathrm{zz}\ \mathrm{[cm^2\ s^{-1}]}$ & Eddy diffusion coefficient & $\mathcal{U}(5.0,15.0)$ & - & $8.70^{+0.25}_{-0.21}$ & - & $8.17^{+0.12}_{-0.09}$ \\
     & & & \underline{Derived$^{\rm (d)}$} & \underline{Derived$^{\rm (d)}$} & \underline{Derived$^{\rm (d)}$} & \underline{Derived$^{\rm (d)}$} \\
    $\mathrm{C/O}$          & $\mathrm{C/O}$ ratio & - & $0.651^{+0.002}_{-0.002}$ & $0.613^{+0.002}_{-0.002}$ & $0.646^{+0.002}_{-0.002}$ & $0.572^{+0.004}_{-0.003}$ \\
    \hline
    $\log\textit{g}\ \mathrm{[cm\ s^{-2}]}$   & Surface gravity & \begin{tabular}{@{}r@{}}$\mathcal{N}_\mathrm{A}(4.96,0.09)$\\$\mathcal{N}_\mathrm{B}(4.88,0.09)$\end{tabular} & $4.96^{+0.02}_{-0.01}$ & $4.96^{+0.02}_{-0.02}$ & $4.73^{+0.02}_{-0.02}$ & $4.82^{+0.01}_{-0.01}$ \\
    $\textit{v}\sin i\ \mathrm{[km\ s^{-1}]}$ & Projected rotation velocity & $\mathcal{U}(10.0,30.0)$ & $14.76^{+0.03}_{-0.03}$ & $14.80^{+0.03}_{-0.03}$ & $24.79^{+0.05}_{-0.04}$ & $25.34^{+0.04}_{-0.04}$ \\
    $\textit{v}_\mathrm{rad}\ \mathrm{[km\ s^{-1}]}$ & Radial velocity & $\mathcal{U}(10.0,30.0)$ & $17.21^{+0.01}_{-0.01}$ & $17.21^{+0.02}_{-0.02}$ & $19.75^{+0.03}_{-0.03}$ & $19.75^{+0.02}_{-0.02}$ \\
    \hline
    $\nabla_1$ & Temperature gradient at $P_1$ & $\mathcal{U}(0.10,0.34)$ & $0.20^{+0.07}_{-0.06}$ & $0.18^{+0.07}_{-0.05}$ & $0.22^{+0.07}_{-0.07}$ & $0.28^{+0.04}_{-0.05}$ \\
    $\nabla_2$ & Temperature gradient at $P_2$ & $\mathcal{U}(0.10,0.34)$ & $0.218^{+0.012}_{-0.013}$ & $0.234^{+0.012}_{-0.010}$ & $0.219^{+0.020}_{-0.022}$ & $0.253^{+0.007}_{-0.007}$ \\
    $\nabla_3$ & Temperature gradient at $P_3$ & $\mathcal{U}(0.05,0.34)$ & $0.051^{+0.001}_{-0.001}$ & $0.052^{+0.002}_{-0.001}$ & $0.091^{+0.002}_{-0.002}$ & $0.057^{+0.004}_{-0.003}$ \\
    $\nabla_4$ & Temperature gradient at $P_4$ & $\mathcal{U}(0.0,0.34)$  & $0.30^{+0.01}_{-0.01}$ & $0.29^{+0.02}_{-0.02}$ & $0.28^{+0.01}_{-0.01}$ & $0.26^{+0.01}_{-0.01}$ \\
    $\nabla_5$ & Temperature gradient at $P_5$ & $\mathcal{U}(0.0,0.34)$  & $0.11^{+0.08}_{-0.07}$ & $0.13^{+0.10}_{-0.08}$ & $0.06^{+0.06}_{-0.04}$ & $0.12^{+0.07}_{-0.06}$ \\
    $\log P_3\ \mathrm{[bar]}$ & Pressure of central knot      & $\mathcal{U}(-1.0,1.0)$  & $-0.35^{+0.02}_{-0.02}$ & $-0.19^{+0.03}_{-0.04}$ & $-0.24^{+0.03}_{-0.03}$ & $-0.19^{+0.02}_{-0.02}$ \\
    $T_3\ \mathrm{[K]}$      & Temperature at central knot   & $\mathcal{U}(900,1900)$ & $1191^{+7}_{-7}$ & $1228^{+10}_{-11}$ & $1149^{+11}_{-10}$ & $1134^{+7}_{-6}$ \\
    $\Delta\log P_{23}\ \mathrm{[bar]}$ & Separation of $P_2$ and $P_3$ & $\mathcal{U}(0.5,2.0)$ & $1.01^{+0.07}_{-0.07}$ & $0.77^{+0.08}_{-0.06}$ & $1.60^{+0.23}_{-0.26}$ & $0.72^{+0.05}_{-0.04}$ \\
    $\Delta\log P_{34}\ \mathrm{[bar]}$ & Separation of $P_3$ and $P_4$ & $\mathcal{U}(0.5,2.0)$   & $0.60^{+0.06}_{-0.05}$ & $0.90^{+0.12}_{-0.12}$ & $0.54^{+0.05}_{-0.03}$ & $0.62^{+0.07}_{-0.05}$ \\
    \hline
    $\log\kappa_{\mathrm{cl},0}\ \mathrm{[cm^2\ g^{-1}]}$ & Cloud-base opacity & $\mathcal{U}(-10.0,3.0)$ & $-1.16^{+0.03}_{-0.03}$ & $-1.11^{+0.03}_{-0.03}$ & $-4.18^{+3.20}_{-3.17}$ & $-1.39^{+0.03}_{-0.03}$ \\
    $\log P_{\mathrm{cl},0}\ \mathrm{[bar]}$ & Cloud-base pressure & $\mathcal{U}(0.5,2.5)$ & $0.52^{+0.01}_{-0.01}$ & $0.52^{+0.02}_{-0.01}$ & $1.58^{+0.53}_{-0.57}$ & $0.52^{+0.01}_{-0.01}$ \\
    $f_\mathrm{sed}$ & Cloud-opacity decay & $\mathcal{U}(1.0,20.0)$ & $1.02^{+0.02}_{-0.01}$ & $1.03^{+0.03}_{-0.02}$ & $10.63^{+5.24}_{-5.40}$ & $1.03^{+0.02}_{-0.01}$ \\
    \hline
    $\log a$    & GP amplitude    & $\mathcal{U}(-0.7,0.5)$ & $0.198^{+0.004}_{-0.004}$ & $0.198^{+0.005}_{-0.005}$ & $0.124^{+0.005}_{-0.005}$ & $0.108^{+0.003}_{-0.003}$ \\
    $\log \ell\ \mathrm{[km\ s^{-1}]}$ & GP length-scale & $\mathcal{U}(-3.0,-1.0)$ & $-1.300^{+0.005}_{-0.001}$ & $-1.300^{+0.005}_{-0.001}$ & $-1.351^{+0.010}_{-0.004}$ & $-1.324^{+0.005}_{-0.004}$ \\
    \hline
     & & $\mathbf{\Delta BIC}$ & $\mathbf{0}$ & $\mathbf{-20.31}$ & $\mathbf{0}$ & $\mathbf{-305.67}$ \\
\end{longtable}
\tablefoot{(a) The disequilibrium mixing ratios are not directly retrieved but instead derived near the photosphere, at $2\ \mathrm{bar}$. (b) The free-chemistry elemental abundances and isotopologue ratios are derived from the retrieved VMRs above. (c) The elemental abundances are projected onto the expected surface gravities, as described in Sect. \ref{sect:elemental_abundances} and shown in brackets. (d) The $\mathrm{C/O}$ ratios derived from the free-chemistry model trace only the gaseous carbon and oxygen, while the disequilibrium model accounts for oxygen condensation.}

\newpage
\section{Best-fitting spectra} \label{app:best_fitting_spectra}
\begin{figure*}[h!]
    \centering
    \includegraphics[width=17cm]{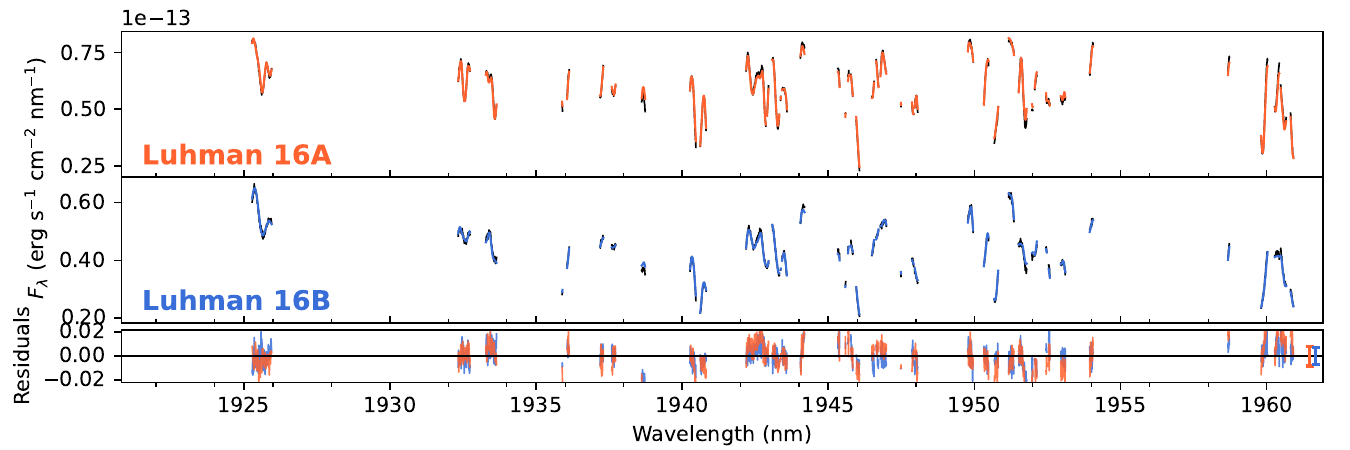}
    \includegraphics[width=17cm]{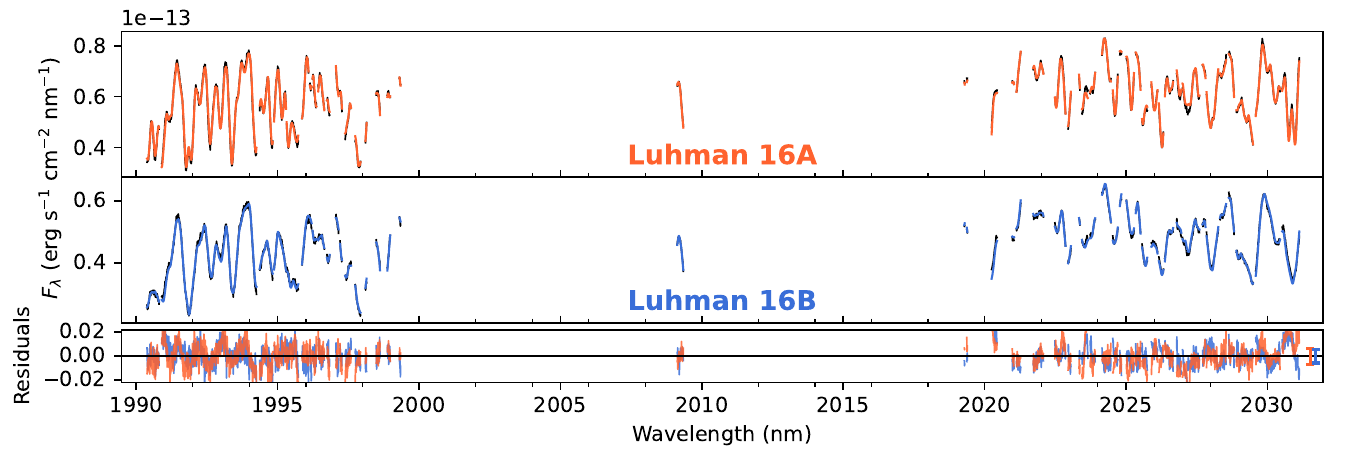}
    \includegraphics[width=17cm]{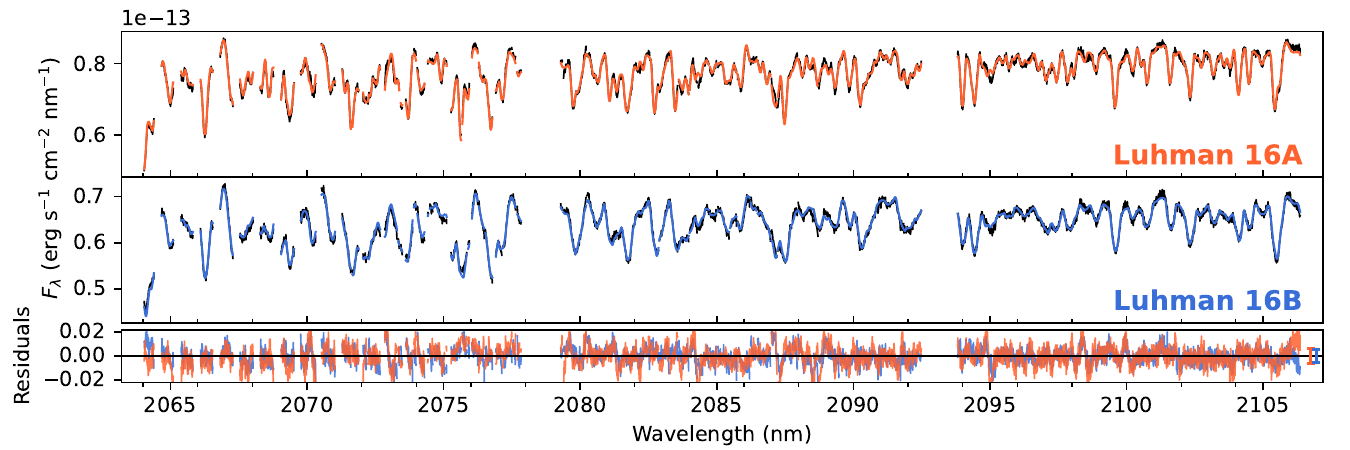}
    \includegraphics[width=17cm]{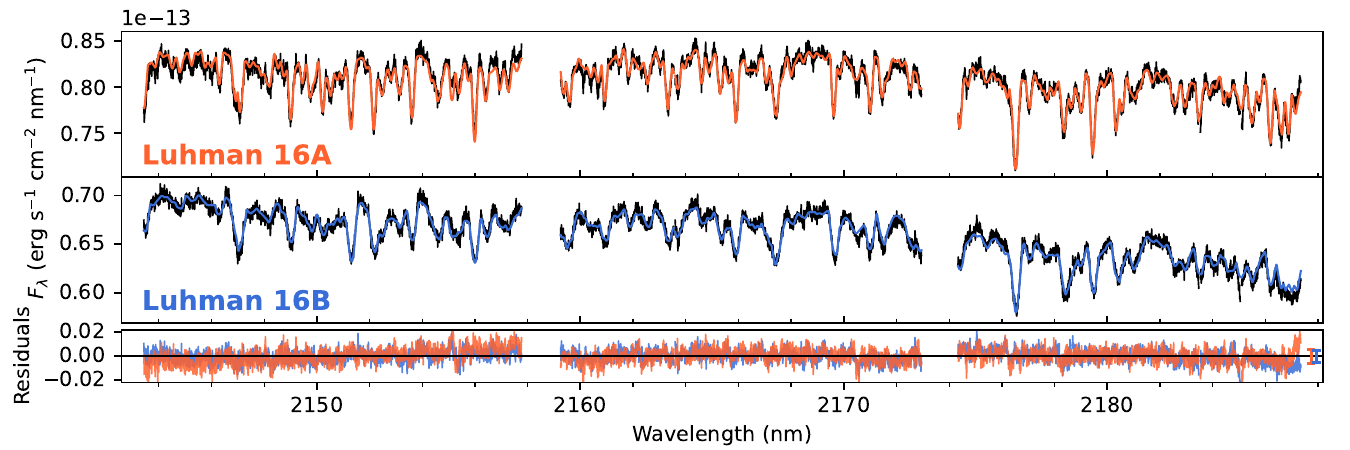}
    \caption{Same as Fig.~\ref{fig:spectrum} but showing all spectral orders.}
    \label{fig:spectrum_app1}
\end{figure*}
\begin{figure*}[h!]
\addtocounter{figure}{-1}
    \centering
    \includegraphics[width=17cm]{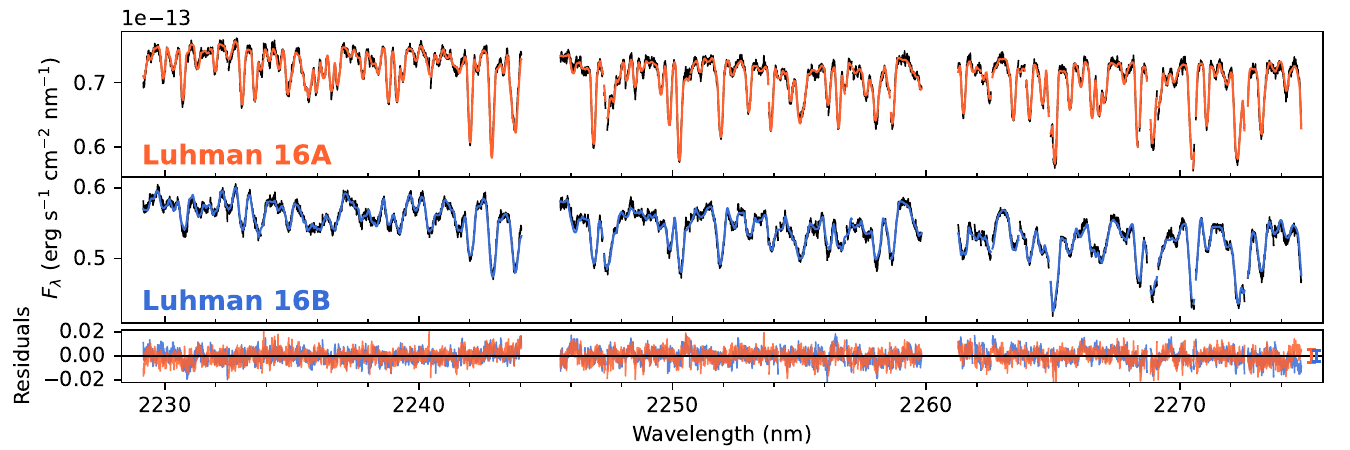}
    \includegraphics[width=17cm]{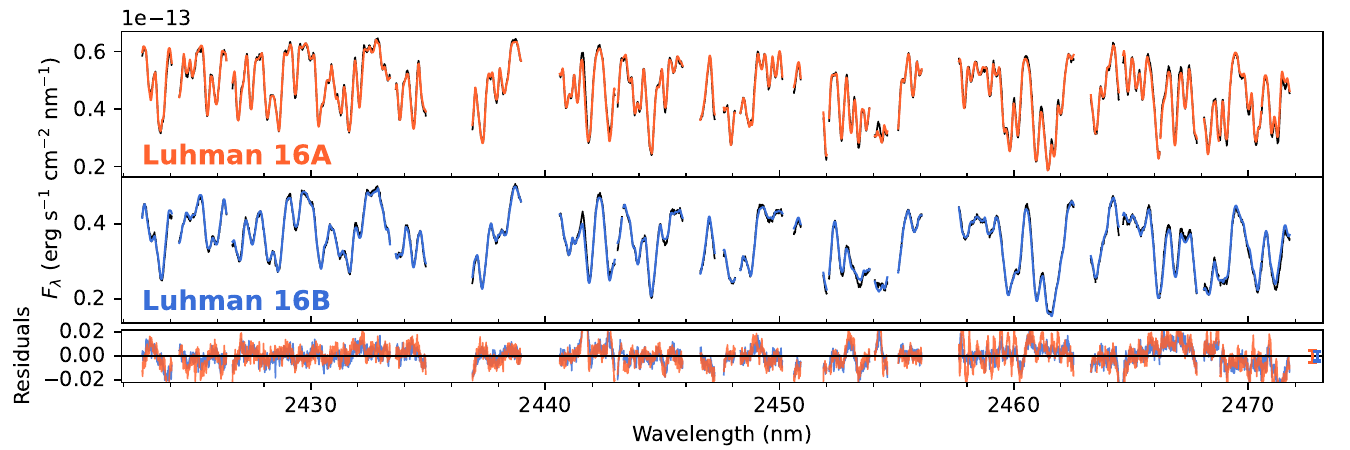}
    \captionsetup{labelformat=empty}
    \caption{Continued.}
\end{figure*}

\end{appendix}

\end{document}